\documentclass[epjc3]{svjour3}  

\usepackage{graphicx}
\usepackage{fix-cm}
\usepackage{float}
\usepackage{afterpage}
\usepackage{epsfig,cite}
\usepackage{amssymb}
\usepackage{amsmath}
\usepackage{dsfont}
\usepackage{multirow}
\usepackage{url}
\usepackage{xcolor}
\usepackage{float}
\usepackage{afterpage}
\usepackage{booktabs}
\usepackage[right=2.5cm]{geometry}

\usepackage{tikz}
\usetikzlibrary{arrows}
\usepackage{tikz-3dplot}
\usepackage{enumitem}

\usepackage{amsfonts}
\usepackage{pifont}
\usepackage{epstopdf}
\usepackage{epsfig}
\usepackage[framed]{ntheorem}
\usepackage{framed}
\usepackage{makeidx}
\usepackage{simplewick}
\usepackage{hyperref}
\usepackage{placeins}
\usepackage{color}
\usepackage{fancyvrb}

\DefineVerbatimEnvironment{Highlighting}{Verbatim}{commandchars=\\\{\}}
\newcommand{\be}{\begin{equation}}
\newcommand{\ee}{\end{equation}}
\newcommand{\bea}{\begin{eqnarray}}
\newcommand{\eea}{\end{eqnarray}}
\newcommand{\bi}{\begin{itemize}}
\newcommand{\ei}{\end{itemize}}
\newcommand{\ben}{\begin{enumerate}}
\newcommand{\een}{\end{enumerate}}
\newcommand{\la}{\left\langle}
\newcommand{\ra}{\right\rangle}

\newcommand{\lp}{\left(}
\newcommand{\rp}{\right)}

\def\frac#1#2{{{#1}\over {#2}}}
\def\gsim{\mathrel{\rlap{\lower4pt\hbox{\hskip1pt$\sim$}}
    \raise1pt\hbox{$>$}}}         %greater than or approx. symbol
\def\lsim{\mathrel{\rlap{\lower4pt\hbox{\hskip1pt$\sim$}}
    \raise1pt\hbox{$<$}}}         %less than or approx. symbol

\newcommand{\draft}[1]{}

\numberwithin{equation}{section}
\numberwithin{figure}{section}
\numberwithin{table}{section}

\usepackage{tabularx}
\newcolumntype{C}[1]{>{\centering\arraybackslash}p{#1}}

\renewcommand{\vec}[1]{\textbf{#1}}

\newcommand{\simunet}{\texttt{SIMUnet}}
\newcommand{\fitmaker}{\texttt{fitmaker}}
\newcommand{\smefit}{\texttt{SMEFiT}}
\newcommand{\nnpdf}{\texttt{NNPDF4.0}}
\newcommand{\nnpdfcode}{\texttt{NNPDF}}
\newcommand{\apfel}{\texttt{apfel}}

\newcommand{\evolventhreefit}{\texttt{evolven3fit}}
\newcommand{\web}{\href{https://hep-pbsp.github.io/SIMUnet/}{\url{https://hep-pbsp.github.io/SIMUnet/}}}

\usepackage{neuralnetwork}
\usepackage{xstring}
\usepackage{xpatch}
\usepackage{array}

\makeatletter
\def\thickhline{%
             \noalign{\ifnum0 =`}\fi\hrule \@height \thickarrayrulewidth \futurelet
             \reserved@a\@xthickhline}
\def\@xthickhline{\ifx\reserved@a\thickhline
                \vskip\doublerulesep
                \vskip -\thickarrayrulewidth
                \fi
                \ifnum0 =`{\fi}}
\xpatchcmd{\linklayers}{\nn@lastnode}{\lastnode}{}{}
\xpatchcmd{\linklayers}{\nn@thisnode}{\thisnode}{}{}
\makeatother
\newlength{\thickarrayrulewidth}
 \makeatletter
 \renewcommand{\section}{%
  \@startsection{section}
    {0}
    {\z@}
    {-21dd plus-8pt minus-8pt}
    {6dd}
    {\normalsize\bfseries\boldmath}%
}
\renewcommand{\subsection}{%
  \@startsection{subsection}
    {1}
    {\z@}
    {-21dd plus-8pt minus-14pt}
    {6dd}
    {\normalsize\bfseries\boldmath}%
}
\makeatother

\usepackage{pythonhighlight}
\journalname{Eur. Phys. J. C}
\begin{document}
\title{\simunet{}: an open-source tool for simultaneous global fits\\ of EFT
  Wilson coefficients and PDFs}
\author{The PBSP collaboration:\\ Mark N. Costantini\thanksref{e0,addr1}
        \and Elie Hammou\thanksref{e1,addr1} \and \\
        Zahari Kassabov\thanksref{addr1} \and Maeve Madigan\thanksref{e2,addr2}
         \and Luca Mantani\thanksref{e3,addr1}
         \and \\ Manuel Morales-Alvarado\thanksref{e4,addr1}
         \and James M. Moore\thanksref{e5,addr1}
         \and \\Maria Ubiali\thanksref{e6,addr1}}

\thankstext{e0}{e-mail: mnc33@cam.ac.uk}
\thankstext{e1}{e-mail: eh651@cam.ac.uk}
\thankstext{e2}{e-mail: madigan@thphys.uni-heidelberg.de}
\thankstext{e3}{e-mail: lm962@cam.ac.uk}
\thankstext{e4}{e-mail: mom28@cam.ac.uk}
\thankstext{e5}{e-mail: jmm232@cam.ac.uk}
\thankstext{e6}{e-mail: M.Ubiali@damtp.cam.ac.uk}

\institute{ DAMTP, University of Cambridge, Wilberforce Road, Cambridge, CB3 0WA, UK \label{addr1} 
\and Institut f\"{u}r Theoretische Physik, Universit\"{a}t Heidelberg, Philosophenweg 16, D-69120, Heidelberg, Germany\label{addr2}}

\date{Received: date / Accepted: date}
\maketitle

\begin{abstract}

We present the open-source \simunet{} code, designed to fit Standard
Model Effective Field Theory (SMEFT) Wilson coefficient alongside
Parton Distribution Functions (PDFs) of the proton.
\simunet{} can perform SMEFT global fits, as well as simultaneous fits
of the PDFs and of an arbitrarily large number of SMEFT degrees of
freedom, by including both PDF-dependent and PDF-independent
observables. \simunet{} can also be used to determine whether the effects of any New Physics 
models can be fitted away in a global fit of PDFs. 
\simunet{} is built upon the open-source \nnpdfcode{} code and is
released together with documentation, and tutorials.
To illustrate the functionalities of the new tool, we present a new
global analysis of the SMEFT Wilson coefficients
accounting for their interplay with the PDFs. We increment our
previous analysis of the LHC Run II top quark data with both (i) the
Higgs production and decay rates data from the LHC, and (ii) the
precision electroweak and diboson measurements from LEP and the LHC.
\end{abstract}

\tableofcontents

\newpage

\section{Introduction}
\label{sec:introduction}

The success of the ambitious programme of the upcoming Run III at the LHC and its subsequent
High-Luminosity (HL-LHC) upgrade~\cite{Cepeda:2019klc,Azzi:2019yne} relies not only on
achieving
the highest possible accuracy in the experimental measurements and in the corresponding
theoretical predictions, but also on the availability of statistically robust tools capable of yielding
global interpretations of all subtle deviations from the Standard Model (SM)
that the data might indicate.
The lack of evidence for additional particles at the LHC or at other colliders
so far suggests that any particles beyond the Standard Model (BSM) may
be significantly heavier than the scale probed by the LHC. Hence, the effects of heavy 
BSM particles may be approximated by integrating them out to obtain higher-dimensional
interactions between the SM fields~\cite{Appelquist:1974tg}. The SM may then be seen as
an effective field theory (EFT) and can be supplemented by higher-dimensional operators
that are suppressed by inverse powers of new particles' mass
scale~\cite{Weinberg:1979sa,Buchmuller:1985jz,Grzadkowski:2010es}.

The Standard Model Effective Field Theory (SMEFT) is a powerful framework to constrain,
identify, and parametrise potential deviations with respect to SM predictions;
see Ref.~\cite{Brivio:2017vri} for a review. It allows for the interpretation of experimental
measurements in the context of BSM scenarios featuring heavy new particles while
minimising assumptions on the nature of the underlying UV-complete theory. 
Early SMEFT analyses constrained subsets of SMEFT operators relevant to sectors 
of observables, for example the electroweak, diboson and Higgs 
sectors~\cite{Han:2004az,daSilvaAlmeida:2018iqo,Ellis:2014jta,Almeida:2021asy,Biekotter:2018ohn,Kraml:2019sis,Ellis:2018gqa,Corbett:2012ja,Ethier:2021ydt},
the top sector~\cite{Buckley:2015lku,Brivio:2019ius,Hartland:2019bjb,Bissmann:2019gfc,Kassabov:2023hbm,Elmer:2023wtr} 
as well as Drell-Yan~\cite{Allwicher:2022gkm,Boughezal:2022nof}.
More recently, global fits constraining a larger set of SMEFT 
operators have been performed: the Higgs, top, diboson and electroweak sectors have 
been combined~\cite{Ellis:2020unq,Ethier:2021bye,Giani:2023gfq}, as well as combinations of those sectors alongside Drell-Yan
and flavour observables~\cite{Bartocci:2023nvp}.

%before
%While early SMEFT analyses included only sector-specific operators, modern SMEFT
%analyses provide global fits in a complete basis of operators within some
%flavour assumptions. FitMaker~\cite{Ellis:2020unq} and 
%SMEFiT~\cite{Giani:2023gfq,Ethier:2021bye,Hartland:2019bjb} ...
%\MU{Some text needed here}.

The determination of SMEFT Wilson coefficients from a fit of LHC data,
like the determination of SM precision parameters from LHC data, might 
display a non-negligible interplay with the input set of Parton Distribution
Functions (PDFs) used to compute theory predictions. This was shown in
several recent
studies~\cite{Carrazza:2019sec,Greljo:2021kvv,Gao:2022srd,Iranipour:2022iak,Kassabov:2023hbm,Hammou:2023heg}, 
in which such interplay between PDFs and SMEFT Wilson
coefficients was quantified for the first time in relevant phenomenological scenarios.
For example, in~\cite{Carrazza:2019sec} it was shown that, while the effect of four-fermion operators
on Deep Inelastic Scattering (DIS) data can be non-negligible, if DIS data were fitted
while taking the effect of such operators into account, the fit quality would deteriorate
proportionally to the energy scale $Q^2$ of the data included in the determination. On the other hand
in the context of high-mass Drell-Yan, especially in the HL-LHC scenario, neglecting the
cross-talk between the large-$x$ PDFs and the SMEFT effects in the
tails could potentially miss new physics manifestations or misinterpret
them~\cite{Greljo:2021kvv,Madigan:2021uho,Hammou:2023heg}, as the bounds on SMEFT operators are significantly
broader if the PDFs are fitted by including the effect of universal operators in the
high energy tails of the data. In~\cite{Hammou:2023heg} it was shown that the large-$x$
antiquark distributions can absorb the effect of universal new physics in the tails of the 
Drell-Yan distributions by leading to significantly softer antiquark distributions in the large-$x$ region, 
far beyond the nominal PDF uncertainties. The analysis of the 
top quark sector of Ref.~\cite{Kassabov:2023hbm} demonstrates that in
that case the bounds of the operators are not broadened 
by the interplay with the PDFs, however the correlation between the top sector and the 
gluon is manifest in the gluon PDF itself, which becomes softer in the large-$x$ region if 
the PDF fit is augmented by the top data and PDFs are fitted simultaneously along the 
Wilson coefficients that determine the top quark sector. 

The first exploration of how the Wilson coefficients of 
the SMEFT and the proton PDFs are intertwined~\cite{Carrazza:2019sec,Greljo:2021kvv,Gao:2022srd} was 
somewhat limited by methodologies that allowed for the consideration of only a handful of SMEFT operators, 
as they were based on a scan in the operator space.
\simunet{} was the first tool that allowed for a 
truly simultaneous fit of the PDFs alongside any physical parameter that enters
theoretical predictions, whether a precision SM parameter, or the Wilson coefficients of
a SMEFT expansion without any hard limits in the number of parameters that are fitted
alongside the PDFs. 
The new framework was first presented in Ref.~\cite{Iranipour:2022iak} in the context
of high-mass Drell-Yan distributions. Later in Ref.~\cite{Kassabov:2023hbm} it was
applied to analyse the broadest top quark dataset used to date in either PDF or SMEFT
interpretations, which in particular contained all available
measurements in the top sector from the ATLAS and CMS experiments based on the full Run II
luminosity.
By combining this wide dataset the potentiality of the \simunet{} methodology was
showcased by using it to derive bounds on 25 independent Wilson coefficients,
both independently and alongside the PDFs.

In this paper, we present and publicly release the computational framework \simunet{}, so that any user 
can assess the interplay between PDFs and New Physics, either by performing simultaneous fits of 
SMEFT operators (or operators in any other EFT expansion) alongside the PDFs of the proton, or by injecting any New Physics model in the 
data and checking whether a global fit of PDFs can absorb the effects induced by such a model in the data.  

\simunet{} is based on the first tagged version of the public {\tt NNPDF} code~\cite{nnpdfcode,NNPDF:2021uiq}
associated with the \nnpdf{} PDF set release~\cite{NNPDF:2021njg,NNPDF:2024dpb,NNPDF:2024djq}, which it augments with
a number of new
features that allow the exploration of the correlation between a fit of PDFs and 
BSM degrees of freedom. First of all, it allows for the simultaneous determination of 
PDFs alongside any Wilson coefficients of an EFT that enters in the theory predictions. The user can 
specify any subset of operators that are of phenomenological relevance, 
compute the corresponding corrections to the SM predictions, and derive 
bounds on the operators from all observables entering the PDF fit and an arbitrary number of 
PDF-independent observables that can be added on top of the PDF-dependent ones. 
Histograms of the bounds on Wilson coefficients, correlation coefficients between PDF and 
Wilson coefficients, as well as Fisher information matrices and bound comparisons are automatically 
produced by the {\tt validphys} analysis framework, a package implemented in
\texttt{reportengine}~\cite{zahari_kassabov_2019_2571601} that allows for the analysis and plotting of data related to the
\simunet{} fit structures and input/output capabilities to other elements of the code base. 
Users can also inject the effects of any New Physics scenario in the data, and assess whether PDFs might 
absorb them and fit them away, in the context of a closure test based on artificial data, 
as carried out in Ref.~\cite{Hammou:2023heg}.

Furthermore, we present a new global simultaneous analysis of SMEFT
Wilson coefficients and PDFs based on a broad range of 
observables, including for the first time  (i) the Higgs production and decay
rates data from the LHC, (ii) precision electroweak and diboson measurements 
from LEP and the LHC, and (iii) Drell-Yan measurements, in a global analysis of 40 dimension-6 SMEFT operators included
linearly. The Wilson coefficients are determined both with fixed PDFs and simultaneously with the PDFs themselves.

The structure of the paper is the following. In Section~\ref{sec:code} we review the \simunet{} 
methodology and describe in detail how it is implemented in the code. This section is 
especially relevant for those who wish to use the public code to perform simultaneous fits of 
PDFs and Wilson coefficients of an EFT. In Section~\ref{sec:contamination} we describe the use of the code
in assessing whether New Physics signals can be absorbed by PDFs. In Section~\ref{sec:data} 
we present all new data implemented for this release, and present the results 
we obtain in the first global SMEFT analysis that explores up to 40 Wilson Coefficients 
and fits them alongside PDFs. We conclude in Section~\ref{sec:conclusions} by providing the links to the 
public code repository and to the online documentation, and outlining the plan for future developments.

\section{\simunet{}: methodology for simultaneous fits and code structure}
\label{sec:code}
In this section, we provide a broad overview of the general methodology and code structure of \simunet{},
in particular describing its functionality regarding simultaneous PDF-SMEFT fits. In Sect.~\ref{sec:nnpdf} 
we begin by briefly reviewing the \nnpdf{} framework and 
code, which \simunet{} builds upon, and then in Sect.~\ref{sec:simunet} we review the specific \simunet{} methodology and code implementation. 
In Sect.~\ref{sec:codedes} we give some details of usage (in particular the user-facing \texttt{yaml} runcard
structure), but the user is encouraged to check the
supporting webpage, \web{}, for a more comprehensive description. 

%We begin with a review of the key features of the \nnpdf{} framework. We then describe how \simunet{} extends
%the \nnpdf{} framework to allow for the simultaneous determination of PDFs and physical parameters, assuming 
%that the theory predictions are linearly dependent on the relevant physical parameters, to within a good approximation. 

\subsection{Review of the \nnpdf{} framework} 
\label{sec:nnpdf}
The central premise of the \nnpdf{} methodology~\cite{NNPDF:2021uiq,NNPDF:2021njg} is the \textit{neural network} parametrisation
of the PDFs of the proton. The \nnpdf{} methodology assumes that one can write the
eight fitted PDF flavours at the parametrisation scale $Q_0^2$
%$ = (1.65\ \text{GeV})^2$
, $\vec{f}(x, Q_0^2) \in \mathbb{R}^8$, in the form:\footnote{Technically, the form $f_i(x,Q_0^2) = x^{\alpha_i} (1-x)^{\beta_i} NN_{i}(x,\pmb{\omega})$ 
is used for each of the fitted PDF flavours, i.e. a power law scaling is used as a pre-factor; we will ignore this to streamline the presentation. 
Indeed, it was recently shown in Ref.~\cite{Carrazza:2021yrg} that this pre-factor can be removed completely, and recovered using only the neural 
network parametrisation as presented in this text. }
\begin{equation}
\vec{f}(x,Q_0^2) = \vec{NN}(x, \pmb{\omega}),
\end{equation}
where $\vec{NN}(\cdot,\pmb{\omega}) : [0,1] \rightarrow \mathbb{R}^8$ denotes a suitable neural network, and $\pmb{\omega}$ are the parameters of the network. 
Given an $N_{\text{dat}}$-dimensional dataset $\vec{D} \in \mathbb{R}^{N_{\text{dat}}}$, the corresponding theory predictions 
$\vec{T}(\pmb{\omega},\vec{c}) \in \mathbb{R}^{N_{\text{dat}}}$ are constructed from this neural network parametrisation via the following 
discretisation of the standard collinear factorisation formula:
\begin{equation}
\label{eq:convolution}
T_i(\pmb{\omega},\vec{c}) = \begin{cases} \displaystyle \sum_{a = 1}^{N_{\text{flav}}} \sum_{\alpha=1}^{N_{\text{grid},i}}  \text{FK}_{i,a\alpha}(\vec{c}) \text{NN}_a(x_\alpha,\pmb{\omega}), & \text{if $D_i$ is deep-inelastic scattering data;} \\[3ex] \displaystyle\sum_{a,b=1}^{N_{\text{flav}}} \sum_{\alpha,\beta=1}^{N_{\text{grid},i}} \text{FK}_{i,a\alpha b\beta}(\vec{c}) \text{NN}_a(x_\alpha,\pmb{\omega}) \text{NN}_b(x_\beta,\pmb{\omega}), & \text{if $D_i$ is hadronic data.}\end{cases}
\end{equation}
Here, $\text{FK}_{i,a\alpha}(\vec{c})$ (or $\text{FK}_{i,a\alpha b \beta}(\vec{c})$) is called the \textit{fast-kernel (FK) table} for the $i$th datapoint, 
which encompasses both the partonic cross section for the process associated to the $i$th datapoint and the coupled evolution of the PDFs from the parametrisation 
scale $Q_0^2$ to the scale $Q^2_i$ associated to the datapoint $i$,  
and $(x_1,...,x_{N_{\text{grid},i}})$ is a discrete $x$-grid for the $i$th datapoint. Note importantly that the theory predictions carry a dependence on 
the parameters of the neural network, $\pmb{\omega}$, but additionally on a vector of physical constants $\vec{c} \in \mathbb{R}^{N_{\text{param}}}$ through 
the FK-tables, such as the strong coupling $\alpha_s(M_z)$, the electroweak parameters in a given electroweak input scheme, the CKM matrix elements, or the masses of the heavy quarks, 
which are fixed to some reference values for any given \nnpdf{} fit. A choice of constants $\vec{c}$ determines a corresponding set of FK-tables, 
which is referred to as a `theory' in the \nnpdf{} parlance.

A PDF fit in the \nnpdf{} framework comprises the determination of the neural network parameters $\pmb{\omega}$, given a choice of 
theory $\vec{c} = \vec{c}^*$, together with an uncertainty estimate for these parameters. This is achieved via the \textit{Monte Carlo replica method}, 
described as follows. Suppose that we are given experimental data $\vec{D}$, together with a covariance matrix $\Sigma$. 
We generate $N_{\text{rep}}$ `pseudodata' vectors, $\vec{D}_1, ..., \vec{D}_{N_{\text{rep}}}$, as samples from the multivariate normal distribution:
\begin{equation}
	\vec{D}_i \sim \mathcal{N}(\vec{D}, \Sigma).
\end{equation}
The default choice for the covariance matrix $\Sigma$ is $\Sigma = \Sigma_{\rm exp}$, i.e. the 
experimental covariance matrix, which includes all information on experimental uncertainties and correlations. 
The latter can also be augmented by a theory covariance matrix $\Sigma_{\rm th}$, hence setting $\Sigma = \Sigma_{\rm exp} + \Sigma_{\rm th}$, where $\Sigma_{\rm th}$ 
includes the effects of nuclear correction uncertainties~\cite{Ball:2018twp,Ball:2020xqw} and missing higher order 
uncertainties~\cite{AbdulKhalek:2019bux,AbdulKhalek:2019ihb,NNPDF:2024dpb} in the theory predictions used in a PDF fit. 
An alternative approach to keep account of theory uncertainties in the fit is to expand the Monte Carlo sampling to the 
space of factorisation and renormalisation scale parameters; this approach was presented in Ref.~\cite{Kassabov:2022orn}. Both options 
can be implemented in the \simunet{} framework, the former in a straightforward way, the latter with some modifications that 
we leave to future releases.  

For each pseudodata $\vec{D}_k$, we find the corresponding best-fit PDF parameters $\pmb{\omega}_k$ by minimising the $\chi_k^2$ loss function, 
defined as a function of $\pmb{\omega}$ as
\begin{equation}
\label{eq:montecarlo_chi2}
\chi^2_k(\pmb{\omega}, \vec{c}^*) = (\vec{D}_k - \vec{T}(\pmb{\omega}, \vec{c}^*))^T \Sigma_{t_0} (\vec{D}_k - \vec{T}(\pmb{\omega}, \vec{c}^*)) \, .
\end{equation}
The minimisation is achieved using \textit{stochastic gradient descent}, which can be applied here because the analytic dependence 
of $\chi^2_k(\pmb{\omega}, \vec{c}^*)$ on $\pmb{\omega}$ is known, since the neural network parametrisation is constructed from 
basic analytic functions as building blocks. Note the following features of $\chi^2_k$:
\begin{itemize}
	\item The $\chi^2$-loss is not computed using the experimental covariance matrix $\Sigma_{\text{exp}}$, but a related object, $\Sigma_{t_0}$, 
  called the $t_0$ covariance matrix. This allows for a faithful propagation of multiplicative uncertainties, as described in Ref.~\cite{Ball:2009qv}.
\item In practice, the $\chi^2$-loss is additionally supplemented by \textit{positivity} and \textit{integrability} penalty terms; these ensure the positivity of observables and the integrability of the PDFs in the small-$x$ region.
\end{itemize}
Given the best-fit parameters $\pmb{\omega}_1, ..., \pmb{\omega}_{N_{\text{rep}}}$ to each of the pseudodata, we now have an ensemble of neural networks 
which determine an ensemble of PDFs, $\vec{f}_1, ..., \vec{f}_{N_{\text{rep}}}$, from which statistical estimators can be evaluated, such as the \textit{mean} 
or \textit{variance} of the initial-scale PDFs. 
%For example, the \textit{mean} or \textit{central} initial-scale PDF is computed as:
%\begin{equation}
%\overline{\vec{f}}(x,Q_0^2) = \frac{1}{N_{\text{rep}}}\sum_{k=1}^{N_{\text{rep}}} \vec{f}_k(x,Q_0^2),
%\end{equation}
%and the \textit{variance} of the initial scale PDFs is computed as:
%\begin{equation}
%\textrm{var}(\vec{f})(x,Q_0^2) = \frac{1}{N_{\text{rep}}} \sum_{k=1}^{N_{\text{rep}}} \vec{f}^2_k(x,Q_0^2) - \overline{\vec{f}}^2(x,Q_0^2),
%\end{equation}
%where the square denotes the element-wise square of the vector $\vec{f}_k$.

\paragraph{Review of the \nnpdf{} code.} 
The methodology sketched above is at the basis of the \nnpdfcode{} public code~\cite{nnpdfcode}. The code comprises 
several packages. To transform the original measurements provided by the experimental 
collaborations, e.g. via {\sc\small HepData}~\cite{Maguire:2017ypu}, 
into a standard format that is tailored for fitting, the \nnpdfcode{} code uses the {\tt buildmaster} {\tt C++} experimental data formatter. 
Physical observables are evaluated as a tensor sum as in Eq.~\eqref{eq:convolution} via the 
{\tt APFELcomb}~\cite{Bertone:2016lga} package that takes hard-scattering partonic matrix element interpolators
from {\tt APPLgrid}~\cite{Carli:2010rw} and {\tt
FastNLO}~\cite{Wobisch:2011ij} (for hadronic processes) and \apfel~\cite{Bertone:2013vaa} (for DIS structure functions) and combines
them with the QCD evolution kernels that evolve the initial-scale PDFs.
 The package also handles NNLO QCD and/or NLO electroweak 
$K$-factors when needed.
The actual fitting code is implemented in the {\tt TensorFlow} framework~\cite{abadi2016tensorflow} via the 
\texttt{n3fit} library. The latter allows for a flexible specification of the neural network model adopted to
parametrise the PDFs, whose settings can be selected automatically via 
the built-in hyperoptimisation tooling ~\cite{2015CS&D....8a4008B}, such as the neural network architecture, 
the activation functions, and the initialisation strategy; the choice of optimiser and
of the hyperparameters related to the implementation in the fit of theoretical constraints such as PDF 
positivity~\cite{Candido:2020yat} and integrability. Finally the code comprises the {\tt validphys} analysis framework, 
which analyses and plots data related to the NNPDF fit structures, and governs input/output capabilities of other elements of the code base.\footnote{The full \nnpdfcode 
code documentation is provided at {\tt \url{https://docs.nnpdf.science}}.}

\subsection{The \simunet{} framework}
\label{sec:simunet}

An important aspect of the \nnpdf{} methodology, as well as of most global PDF analyses, is that
the physical parameters $\vec{c}$ must be chosen and fixed before a PDF fit, so that the resulting PDFs are produced \textit{under
the assumption} of a given theory $\vec{c} = \vec{c}^*$. It is desirable (and in fact necessary in some scenarios; see for example 
Sect. 5.3 of Ref.~\cite{Greljo:2021kvv}) to 
relax this requirement, and instead to fit both the PDFs \textit{and} the physical parameters $\vec{c}$ \textit{simultaneously}.

On regarding Eq.~\eqref{eq:montecarlo_chi2}, one may assume that this problem can immediately 
be solved by applying stochastic gradient descent not only to the parameters of the neural network $\pmb{\omega}$, but to the tuple 
$(\pmb{\omega}, \vec{c})$, without first fixing a reference value $\vec{c} = \vec{c}^*$. Unfortunately, this is not a solution in practice; 
the FK-tables typically have an extremely complex, non-analytical, dependence on the physical parameters $\vec{c}$, and evaluation of a 
complete collection of FK-tables at even one point $\vec{c} = \vec{c}^*$ requires hundreds of computational hours and an extensive suite of codes.
Hence, since the dependence of the FK-tables as an analytic function of $\vec{c}$ is unavailable, it follows that $\vec{T}(\pmb{\omega},\vec{c})$ 
is not a known analytic function of $\vec{c}$, and stochastic gradient descent \textit{cannot} be applied.

Thus, if we would like to continue with our programme of simultaneous extraction of PDFs and physical parameters, we will need to 
approximate. The central conceit of the \simunet{} methodology is the observation that for many phenomenologically-interesting 
parameters, particularly the Wilson coefficients of the SMEFT expansion, the dependence of the theory predictions $\vec{T}(\pmb{\omega}, \vec{c})$ on the 
physical parameters $\vec{c}$ can be well-approximated using the linear ansatz:\footnote{Note that in 
Sect.~5.3 of Ref.~\cite{Iranipour:2022iak}, it was initially proposed that \simunet{} 
would operate in \textit{non-linear} regimes too.
However, due to the findings regarding uncertainty propagation using the Monte Carlo replica method presented in 
App. E of Ref.~\cite{Kassabov:2023hbm}, the non-linear feature has been deprecated in this public release of the code.}
\begin{equation}
\label{eq:simunet_predictions}
\vec{T}(\pmb{\omega}, \vec{c}) \approx \vec{T}(\pmb{\omega}, \vec{c}^*) \odot \left[ \vec{1} + \vec{K}_{\text{fac}}(\pmb{\omega}^*) (\vec{c} - \vec{c}^*) \right],
\end{equation}
where $\odot$ denotes the element-wise Hadamard product of vectors,\footnote{Recall that if $\vec{v} = \vec{w} \odot \vec{z}$, the elements of $\vec{v}$ are 
defined by $v_i = w_i z_i$.} $\vec{1} \in \mathbb{R}^{N_{\text{dat}}}$ is a vector of ones, $\vec{c}^*$ is some reference value of the physical parameters, 
and $\vec{K}_{\text{fac}}(\pmb{\omega}^*) \in \mathbb{R}^{N_{\text{dat}} \times N_{\text{param}}}$ is a matrix of pre-computed `$K$-factors':
\begin{equation}
K_{\text{fac}}(\pmb{\omega}^*)_{ij} = \frac{T_i(\pmb{\omega}^*, c_1^*, ..., c_{j-1}^*,c_j^* + 1,c_{j+1}^*, ..., c_{N_{\text{param}}}^*) - T_i(\pmb{\omega}^*,\vec{c}^*)}{T_i(\pmb{\omega}^*, \vec{c}^*)},
\end{equation}
which are designed to approximate the appropriate (normalised) gradients:
\begin{equation}
\label{eq:gradient_approximation}
K_{\text{fac}}(\pmb{\omega}^*)_{ij} \approx \frac{1}{T_i(\pmb{\omega}^*, \vec{c}^*)} \frac{\partial T_i}{\partial c_j}(\pmb{\omega}^*, \vec{c}^*).
\end{equation}
Observe that the $K$-factors are determined with a fixed choice of reference PDF, $\pmb{\omega} = \pmb{\omega}^*$, where they should technically 
depend on PDFs freely. In practice however, this approximation is often justified, and the reliability of the approximation can always be checked post-fit; 
see Appendix C of Ref.~\cite{Greljo:2021kvv} for an example of this validation.

Note that the dependence of the theory in Eq.~\eqref{eq:simunet_predictions} on $\pmb{\omega}$ \textit{and} $\vec{c}$ is now known as an analytic function. 
The \simunet{} methodology, at its heart, now simply extends the \nnpdf{} methodology by replacing the theory predictions in Eq.~\eqref{eq:montecarlo_chi2} 
with those specified by Eq.~\eqref{eq:simunet_predictions}, and then running stochastic gradient descent on a series of Monte Carlo pseudodata as in the 
\nnpdf{} framework. The result of the fit is a series of best-fit tuples 
$(\pmb{\omega}_1, \vec{c}_1)$, ..., $(\pmb{\omega}_{N_{\text{rep}}}, \vec{c}_{N_{\text{rep}}})$ to the various pseudodata, from which statistical estimators 
can be calculated as above.

\subsection{Structure and usage of the \simunet{} code}
\label{sec:codedes} 

\begin{figure}[t]
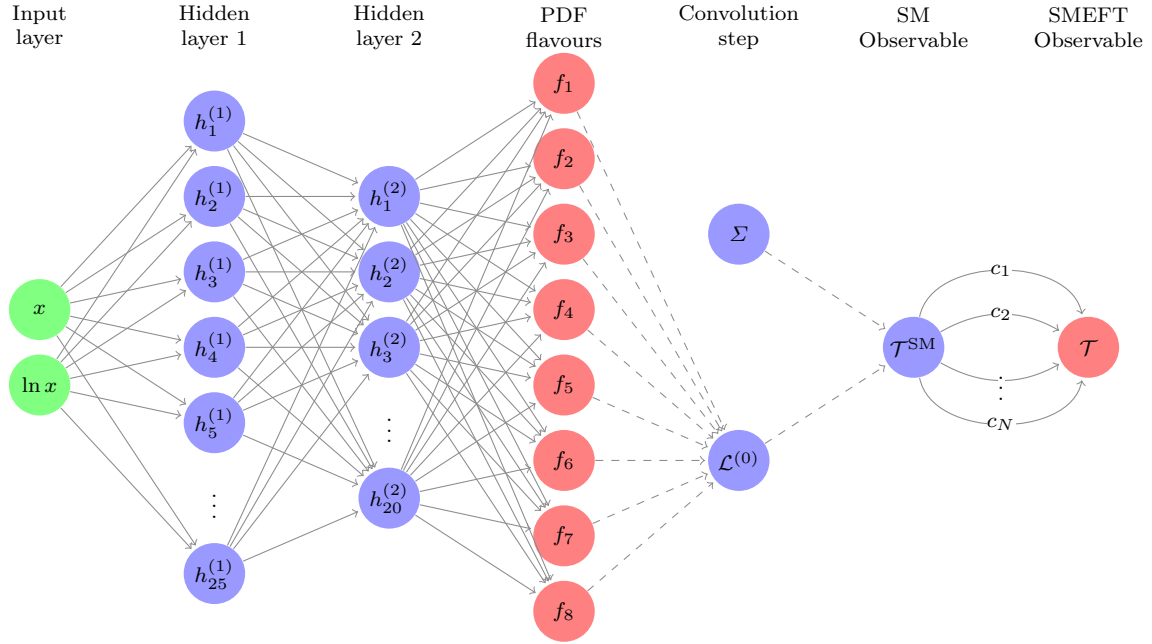

  \centering
  \begin{neuralnetwork}[height=8, layerspacing=23mm, nodesize=23pt]
    \newcommand{\x}[2]{\IfEqCase{#2}{{1}{\small $x$}{2}{\small $\ln{x}$}}}
    \newcommand{\yy}[2]{$f_#2$}
    \newcommand{\basis}[2]{\ifnum #2=4 \(\mathcal{L}^{(0)}\) \else $\Sigma$ \fi}
    \newcommand{\hfirst}[2]{\ifnum #2=7 $h^{(1)}_{25}$ \else $h^{(1)}_#2$ \fi}
    \newcommand{\hsecond}[2]{\ifnum #2=5 $h^{(2)}_{20}$ \else $h^{(2)}_#2$ \fi}
    \newcommand{\standardmodel}[2] {$\mathcal{T}^\text{SM}$}
    \newcommand{\eft}[2] {$\mathcal{T}$}
    \newcommand{\co}[4]{$c_1$}
    \newcommand{\ct}[4]{$c_2$}
    \newcommand{\cnn}[4]{$c_{N-1}$}
    \newcommand{\cn}[4]{$c_{N}$}
    \newcommand{\vd}[4]{$\vdots$}
    \newcommand{\FK}[2]{$\sigma$}
    \newcommand{\convolve}[4]{$\otimes$}
    \newcommand{\ci}[4]{$c_1$}
    \newcommand{\cj}[4]{$c_2$}
    \newcommand{\ck}[4]{$c_N$}
    \inputlayer[count=2, bias=false, title=Input\\layer, text=\x]
    \hiddenlayer[count=7, bias=false, title=Hidden\\layer 1, text=\hfirst, exclude={6}]\linklayers[not to={6}]
    \hiddenlayer[count=5, bias=false, title=Hidden\\layer 2, text=\hsecond, exclude={4}]\linklayers[not from={6}, not to={4}]
    \outputlayer[count=8, title=PDF\\flavours, text=\yy] \linklayers[not from={4}]
    \hiddenlayer[count=4, bias=false, text=\basis, title=Convolution\\step, exclude={2,3}]\linklayers[not to={1,2,3}, style={dashed}]
    \hiddenlayer[count=1, bias=false, title=SM\\Observable, text=\standardmodel]\linklayers[not from={2,3}, style={dashed}]
    \outputlayer[count=1, bias=false, text=\eft, title=SMEFT \\ Observable]
    \link[from layer = 5, to layer = 6, from node = 1, to node = 1, style={bend left=79}, label={\ci}]
    \link[from layer = 5, to layer = 6, from node = 1, to node = 1, style={bend left=30}, label={\cj}]
    \link[from layer = 5, to layer = 6, from node = 1, to node = 1, style={bend right=30}, label={\vd}]
    \link[from layer = 5, to layer = 6, from node = 1, to node = 1, style={bend right=79}, label={\ck}]
    \path (L1-5) -- node{$\vdots$} (L1-7);
    \path (L2-3) -- node{$\vdots$} (L2-5);
  \end{neuralnetwork}
\caption{\label{fig:simunet_network} A schematic representation of the neural network parametrisation of $\vec{T}(\pmb{\omega}, \vec{c})$ used by the \simunet{} code. The usual NNPDF network is represented by the layers from the `Input layer' to the `SM Observable' layer. The final `SMEFT observable' layer is related to the `SM observable' layer by edges which carry the theory parameters $\vec{c}$ as weights.}
\end{figure}

The \simunet{} code is a fork of the \nnpdf{} public code~\cite{NNPDF:2021uiq}, where the key modification is the replacement of the standard theory 
predictions used in the $\chi^2$-loss, Eq.~\eqref{eq:montecarlo_chi2}, by \nnpdf{} with the approximate formula for the theory predictions given 
in Eq.~\eqref{eq:simunet_predictions}. This replacement is effected by the inclusion of a new post-observable \textit{combination layer} in the network, 
specified by the \texttt{CombineCfac.py} layer added to the \texttt{n3fit/layers} directory; see Fig.~\ref{fig:simunet_network} for a schematic representation.

Beyond the inclusion of this layer, the main modification to the \nnpdf{} code comprises reading of the $K$-factors required in the approximation shown in Eq.~\eqref{eq:simunet_predictions}. The $K$-factors are implemented in a new file format, the \texttt{simu\_fac} format. One \texttt{simu\_fac} file is made available for each dataset. The \texttt{simu\_fac} files are packaged inside the relevant \nnpdf{} theory folders, inside the \texttt{simu\_factors} directory; an example of the correct structure is given inside \texttt{theory\_270}, which is a theory available as a separate download to the \simunet{} release. As an example of the files, consider the file \texttt{SIMU\_ATLAS\_CMS\_WHEL\_8TEV.yaml} from the directory \texttt{theory\_270/simu\_factors}:
\begin{python}
metadata:
  ref: arXiv:2005.03799
  author: Luca Mantani
  ufomodel: SMEFTatNLO
  flavour: dim6top
  date: 15/09/2023
  pdf: None
  info_SM_fixed: SMEFiT
  observable: F0, FL
  bins: []

SM_fixed: [0.6896346, 0.3121927]

EFT_LO:
  SM: [0.697539, 0.30189]
  OtW: [-0.0619917432825, 0.0620585212469]
  OtG: [0.0, 0.0]

EFT_NLO:
  SM: [0.689466, 0.308842]
  OtW: [-0.0609861823272, 0.061150870174]
  OtG: [0.00064474891454, -0.000673970936802]
\end{python}
The structure of the \texttt{simu\_fac} file is as follows:
\begin{itemize}
\item There is a \texttt{metadata} namespace at the beginning of the file, containing information on the file. The metadata is never read by the code, 
and is only included as a convenience to the user. 
\item The \texttt{SM\_fixed} namespace provides the best available SM predictions; these may be calculated at next-to-leading order or 
next-to-next-to-leading order in QCD, including or excluding EW corrections, depending on the process.
\item The remaining parts of the file contain the information used to construct $K$-factors for different models. In this file, two models are included: the SMEFT at leading order in QCD, and the SMEFT at next-to-leading order in QCD.
\end{itemize}
In order to perform a simultaneous fit of PDFs and the parameters specified in a particular model of a \texttt{simu\_fac} file, we must create a runcard.  The runcard
follows the standard \nnpdf{} format, details and examples of which are given on the website \web{}.  We provide here a discussion of
the modifications necessary for a \simunet{} fit.
First, we must modify the \texttt{dataset\_inputs} configuration key. Here is an example of an appropriate \texttt{dataset\_inputs} for a simultaneous fit of PDFs and SMEFT Wilson coefficients at next-to-leading order in QCD:
\begin{python}
dataset_inputs:
- {dataset: NMC, frac: 0.75}
- {dataset: ATLASTTBARTOT7TEV, cfac: [QCD], simu_fac: "EFT_NLO"}
- {dataset: CMS_SINGLETOPW_8TEV_TOTAL, simu_fac: "EFT_NLO", use_fixed_predictions: True}
\end{python}
Here, we are fitting three datasets: the fixed-target DIS data from the New Muon Collaboration~\cite{NewMuon:1996fwh,NewMuon:1996uwk} (\texttt{NMC}), 
the total $t\bar{t}$ cross section measured by ATLAS at $\sqrt{s}=7$~TeV~\cite{ATLAS:2014nxi} (\texttt{ATLASTTBARTOT7TEV}) and 
the total associated single top and $W$ boson cross section measure by CMS at $\sqrt{s}=8$~TeV~\cite{CMS:2014fut} (\texttt{CMS\_SINGLETOPW\_8TEV\_TOTAL}), 
which we discuss in turn:
\begin{enumerate}
\item First, note that the dataset \texttt{NMC} is entered exactly as it would appear in an \nnpdf{} runcard; 
it is therefore treated by \simunet{} as a standard dataset for which there is no $K$-factor modification, 
i.e. it depends purely on the PDFs and no additional physics parameters. 
\item On the other hand, the dataset \texttt{ATLASTTBARTOT7TEV} has an additional key included, namely \texttt{simu\_fac}, 
set to the value \texttt{EFT\_NLO}; this tells \simunet{} to treat theory predictions for this dataset using the approximation 
Eq.~\eqref{eq:simunet_predictions}, with the $K$-factors constructed from the \texttt{EFT\_NLO} model specified in the relevant 
\texttt{simu\_fac} file. Precisely \textit{which} parameters from the model are fitted is determined by the \texttt{simu\_parameters} 
configuration key, which we describe below.
\item Finally, observe that the dataset \texttt{CMS\_SINGLETOPW\_8TEV\_TOTAL} includes two new keys: \texttt{simu\_fac} and 
\texttt{use\_fixed\_predictions}. The first key, \texttt{simu\_fac} has exactly the same interpretation as with the previous dataset; 
on the other hand, the fact that the key \texttt{use\_fixed\_predictions} is set to \texttt{True} instead removes the PDF-dependence of this dataset. 
That is, the predictions for this dataset are simplified even from Eq.~\eqref{eq:simunet_predictions}, to:
\begin{equation}
\label{eq:simunet_predictions_fixed}
\vec{T}(\pmb{\omega}, \vec{c}) \approx \vec{T}_{\text{fixed}} \odot \left[ \vec{1} + \vec{K}_{\text{fac}}(\pmb{\omega}^*) (\vec{c} - \vec{c}^*) \right],
\end{equation}
where $\vec{T}_{\text{fixed}}$ is the vector of predictions taken from the \texttt{SM\_fixed} namespace of the \texttt{simu\_fac} file. Note that the right 
hand side no longer has a dependence on $\pmb{\omega}$, so this dataset effectively becomes PDF-independent. This is appropriate for observables which do not 
depend on the PDFs (e.g. the electroweak coupling, or $W$-helicities in top decays), but also observables which depend only weakly on the PDFs and for which 
the full computation of FK-tables would be computationally expensive.
\end{enumerate}

Outside of the \texttt{dataset\_inputs} configuration key, we must also include a new key, called \texttt{simu\_parameters}. This 
specifies \textit{which} of the parameters in the model read from the \texttt{simu\_fac} files will be fitted, and additionally specifies hyperparameters 
relevant to their fit. An example is:
\begin{python}
simu_parameters:
- {name: 'OtG', scale: 0.01, initialisation: {type: uniform, minval: -1, maxval: 1}}
- {name: 'Opt', scale: 0.1, initialisation: {type: gaussian, mean: 0, std_dev: 1}}
\end{python}
This specification tells \simunet{} that for each of the datasets which have set a model value for \texttt{simu\_fac}, the relevant model in the \texttt{simu\_fac} file should be checked for the existence of each of the parameters, and their contribution should be included if they appear in the model in the file. Observe the following:

\begin{enumerate}
\item The \texttt{scale} can be used to modify the learning rate in the direction of each of the specified parameters. In detail, suppose that the 
scale $\lambda$ is chosen for the parameter $c$. Then, \simunet{} multiplies the relevant $K$-factor contribution by $1/\lambda$ so that we are effectively 
fitting the parameter $\lambda c$ instead of $c$ itself. When $K$-factor contributions from parameters are particularly large, setting a large scale can improve 
training of the network, avoiding exploring parts of the parameter space which have extremely poor $\chi^2$s; see Ref.~\cite{Iranipour:2022iak} for 
further discussion. On the other hand, setting a small scale can speed up training of the network by increasing the effective step size in a particular 
direction in parameter space; the user must tune these scales by hand to obtain optimal results. An automatic scale choice feature may be included in a future release.
\item The \texttt{initialisation} key informs \simunet{} how to initialise the parameters when training commences; this initialisation is random and there 
are three basic types available. If \texttt{uniform} is selected, the initial value of the parameter is drawn from a random uniform distribution between the 
keys \texttt{minval} and \texttt{maxval}, which must be additionally specified. If \texttt{gaussian} is selected, the initial value of the parameter is drawn 
from a random Gaussian distribution with mean and standard deviation given by the keys \texttt{mean} and \texttt{std\_dev} respectively. Finally, 
if \texttt{constant} is chosen, the key \texttt{value} is supplied and the initial value of the parameter is always set to this key (no random selection 
is made in this instance).
\end{enumerate}

It is also possible to specify \textit{linear combinations} of the model parameters to fit; this feature is useful because the supplied \texttt{simu\_fac} 
files only contain SMEFT models with Wilson coefficients in the \textit{Warsaw basis}~\cite{Grzadkowski:2010es}. For example, the user may want to fit in a 
different basis to the 
Warsaw basis, or may wish to fit a UV model which has been matched to the SMEFT at dimension $6$, with parameters given by combinations of the SMEFT 
parameters. An example is the following:
\begin{python}
simu_parameters:
  - name: 'W'
    linear_combination: 
      'Olq3': -15.94
    scale: 1000
    initialisation: {type: uniform, minval: -1, maxval: 1}

  - name: 'Y'
    linear_combination: 
      'Olq1': 1.51606
      'Oed': -6.0606
      'Oeu': 12.1394
      'Olu': 6.0606
      'Old': -3.0394
      'Oqe': 3.0394
    scale: 1000
    initialisation: {type: uniform, minval: -1, maxval: 1}
\end{python}
using the $\hat{W}$ and $\hat{Y}$ parameters studied in Ref.~\cite{Greljo:2021kvv}. Here, we fit the user-defined operators:
\begin{align*}
  \hat{W} &= -15.94 \;\mathcal{O}_{lq}^{(3)}, \\
  \hat{Y} &= 1.51606 \; \mathcal{O}_{lq}^{(1)} - 6.0606 \; \mathcal{O}_{ed} + 12.1394 \; \mathcal{O}_{eu} + 6.0606 \; \mathcal{O}_{lu} - 3.0394 \; \mathcal{O}_{ld} + 3.0394 \; \mathcal{O}_{qe},
\end{align*}
which are specified as linear combinations of SMEFT operators, the contributions of which are supplied in the \texttt{simu\_fac} files.

Once a runcard is prepared, the user simply follows the standard \nnpdf{} pipeline to perform a fit. In particular, they should run \texttt{vp-setupfit}, 
\texttt{n3fit}, \evolventhreefit{} and \texttt{postfit} in sequence in order to obtain results. Analysis of the results can be achieved with a range of tools supplied with the release; further details are given below and on the website: \web{}.

\paragraph{The fixed PDF feature of \simunet{}.} The \simunet{} release also provides functionality for fits of the physical parameters $\vec{c}$ alone, with the PDFs kept fixed (similar to tools such as \smefit{} or \fitmaker{} in the case of the SMEFT Wilson coefficients). This is achieved simply by loading the weights of a previous neural network PDF fit~\footnote{See the documentation in \web{} for a list of available fits to preload.}, then keeping these weights fixed as the remaining parameters are fitted.

To specify this at the level of a \simunet{} runcard, use the syntax:
\begin{python}
fixed_pdf_fit: True
load_weights_from_fit: 221103-jmm-no_top_1000_iterated
\end{python}
for example. Setting \texttt{fixed\_pdf\_fit} to \texttt{True} instructs \simunet{} to perform the fit keeping the weights of the PDF part of the network constant, and the namespace \texttt{load\_weights\_from\_fit} tells \simunet{} which previous PDF fit to obtain the frozen weights from. The pipeline for a fit then proceeds exactly as in the case of a normal \simunet{} fit, beginning by running \texttt{vp-setupfit} and then \texttt{n3fit}. \textit{However}, the user need not use \evolventhreefit{} at the subsequent stage, since the PDF grid is already stored and does not need to be recomputed; it does still need to be copied into the correct fit result directory though, which should be achieved using the \texttt{vp-fakeevolve} script via:
\begin{python}
vp-fakeevolve fixed_simunet_fit num_reps
\end{python}
where \texttt{fixed\_simunet\_fit} is the name of the simultaneous fit that the user has just run, and \texttt{num\_reps} is the number of replicas in the fit. The user \textit{must} still run \texttt{postfit} after running the \texttt{vp-fakeevolve} script.

The ability to load weights from a previous fit can also be used to aid a simultaneous fit, by starting training from a good PDF solution already. For example, if we use the syntax:
\begin{python}
fixed_pdf_fit: False
load_weights_from_fit: 221103-jmm-no_top_1000_iterated
\end{python}
in a \simunet{} runcard, then the weights of the PDF part of the neural network will be initialised to the weights of the appropriate replica from the previous PDF fit \texttt{221103-jmm-no\_top\_1000\_iterated}. \textit{However}, since the namespace \texttt{fixed\_pdf\_fit} is set to \texttt{False} here, the fit will \textit{still} be simultaneous, fitting both PDFs and physical parameters. Assuming that the resulting simultaneously-determined PDF fit is reasonably close to the previous PDF fit, this can significantly decrease training time.

\paragraph{\simunet{} analysis tools.} The \simunet{} code is released with a full suite of analysis tools, 
which build on the tools already available in the \nnpdfcode{} public code. 
These tools rely on the \texttt{validphys} analysis framework implemented in 
\texttt{reportengine}, and address exclusively the PDF aspect of the fits.
They allow the user, for example, to generate PDF plots, luminosities, compare fits, and evaluate fit quality metrics, 
among many other things.

The \simunet{} code adds an extensive set of tools to study results in the SMEFT and assess the PDF-SMEFT interplay.
These analysis tools are exclusively allocated in the \texttt{simunet\_analysis.py} file, where the user can find 
documented functions to perform different tasks. 
In the context of EFT coefficients, \simunet{} calculates, among other things, their posterior distributions, Wilson coefficient bounds, 
correlations and pulls from the SM. 
We refer the reader to the new results presented in Section~\ref{sec:data} to visualise the types of analysis that 
\simunet{} can perform.

\section{Closure tests and `contaminated' fits within \simunet{}}
\label{sec:contamination}
The \nnpdf{} \textit{closure test} methodology, first described in Ref.~\cite{NNPDF:2014otw} and described in much 
more detail in Ref.~\cite{DelDebbio:2021whr}, is regularly deployed by the NNPDF collaboration to ensure that their 
methodology is working reliably. The \simunet{} code extends the capacity of the NNPDF closure test to 
run \textit{contaminated} fits (as performed by the private code used for Ref.~\cite{Hammou:2023heg}) and closure 
tests probing the fit quality of \textit{both} PDFs and physical parameters.

In Sect.~\ref{sec:ct} we begin with a brief review of the \nnpdf{} methodology and identify the key assumption 
that \simunet{} relaxes, namely that the values of the physical parameters used to produce the artificial data are the 
same as those used in the subsequent PDF fit of the artificial data. We then proceed to describe in Sect.~\ref{sec:ctusage}
how to produce the so-called \textit{contaminated} fits, {\it i.e.} fits in which a New Physics model is injected in the 
artificial data, while the fit is done assuming the Standard Model. With this functionality, any users can test 
the robustness of any New Physics scenario against being fitted away in a PDF fit. We give usage details for 
running such fits using the \simunet{} code. 
Finally, in Sect.~\ref{sec:ctall} we discuss how \simunet{} can perform closure tests for both PDFs and physical parameters, 
giving surety of the reliability of the \simunet{} methodology.

\subsection{Review of the NNPDF closure test methodology} 
\label{sec:ct}
The NNPDF closure test methodology begins by supposing that we are given Nature's true PDFs at the initial 
scale $\vec{f}_{\text{true}}(x,Q_0^2) = \vec{NN}(x, \pmb{\omega}_{\text{true}})$,\footnote{Assuming that the PDFs can be 
parametrised by a neural network, and then subsequently using a neural network to perform the fit, ignores modelling error; 
we do not consider this here for simplicity, but both the \nnpdf{} and \simunet{} frameworks can account for it.} and true physical 
parameters $\vec{c}_{\text{true}}$. Under this assumption, experimental data $\vec{D}$ should appear to be a sample from the multivariate 
normal distribution:
\begin{equation}
\vec{D} \sim \mathcal{N}(\vec{T}(\pmb{\omega}_{\text{true}}, \vec{c}_{\text{true}}), \Sigma_{\text{exp}}),
\end{equation}
where $\Sigma_{\text{exp}}$ is the experimental covariance matrix, and $\vec{T}$ is the theory prediction from the full discrete 
convolution formula Eq.~\eqref{eq:convolution} (\textit{not} the \simunet{} prediction Eq.~\eqref{eq:simunet_predictions}). 
A sample from this distribution is called a set of \textit{level 1 pseudodata} in the NNPDF closure test language. Given a set of level 
1 pseudodata, we can attempt to recover the theory of Nature using the \nnpdf{} methodology. Instead of fitting PDFs to experimental data, 
we can perform the fit replacing the experimental data with the level 1 pseudodata. If the resulting PDF fit has a spread covering the true 
PDF law $\vec{f}_{\text{true}}$, we say that the methodology has passed this `closure test'.\footnote{In practice, an ensemble of fits is produced, and if their 68\% confidence bands cover the true law in 68\% of the fits, we say that the methodology has passed the closure test.}

Importantly, closure tests are typically performed \textit{only} fitting the true PDF law; in particular, the closure test usually assumes 
that the true physical parameters $\vec{c}_{\text{true}}$ are known exactly and perfectly, that is, theory predictions for the fit take 
the form $\vec{T}(\pmb{\omega}, \vec{c}_{\text{true}})$ where the PDF parameters $\pmb{\omega}$ are variable and fitted to the pseudodata, but the physical parameters $\vec{c}_{\text{true}}$ are fixed. With \simunet{}, more options become available, as described below.

\subsection{BSM-contaminated fits with \simunet{}} 
\label{sec:ctusage}
In Ref.~\cite{Hammou:2023heg}, closure tests were performed using the standard closure test methodology, but introducing 
a mismatch between the theory used as the `true' underlying law of Nature and the theory used in the fit; 
%using the \textit{wrong} fixed theory _{\text{wrong}}
that is, theory predictions for the closure test fit took the form $\vec{T}(\pmb{\omega}, \vec{c}^*)$ 
where $\vec{c}^* \neq \vec{c}_{\text{true}}$. The authors of Ref.~\cite{Hammou:2023heg} call the result of such a fit 
a \textit{contaminated} fit; 
in that work, this is applied particularly to the case that the fake data is generated with New Physics beyond the Standard 
Model, but the PDFs are fitted 
assuming the Standard Model, i.e. $\vec{c}_{\text{true}}$ is a vector of BSM parameters, and $\vec{c}^*$ is the corresponding vector
where these BSM parameters are set to the values they would take in the Standard Model. In the case that the contaminated fit quality is not noticeably poor, the PDFs are said 
to \textit{reabsorb} New Physics, and 
it is shown in Ref.~\cite{Hammou:2023heg} that this can lead to dangerous consequences in subsequent searches for New Physics.

The \simunet{} code provides the capacity to easily perform contaminated fits, assuming that the underlying theory for the 
level 1 pseudodata is generated 
according to the linear $K$-factor form given by Eq.~\eqref{eq:simunet_predictions}, i.e. the level-1 pseudodata is generated 
as a sample from the multivariate normal distribution with mean:
\begin{equation}
\label{eq:simunet_contaminated}
\vec{T}(\pmb{\omega}_{\text{true}}, \vec{c}^*) \odot \left[ \vec{1} + \vec{K}_{\text{fac}}(\pmb{\omega}^*) (\vec{c}_{\text{true}} - \vec{c}^*) \right],
\end{equation}
where $\pmb{\omega}^*, \vec{c}^*$ are the reference values of the PDF parameters and physical parameters assumed in the relevant model of 
the relevant 
\texttt{simu\_fac} files. This is effected in the \simunet{} code simply by interceding in the standard \nnpdf{} generation of the level 1 
pseudodata, by replacing the central values to which Gaussian noise are added by those presented in Eq.~\eqref{eq:simunet_contaminated}.

Note that the code can be easily modified so that any New Physics can be injected in the level 1 pseudodata, 
without relying on the SMEFT expansion, but simply by computing the multiplicative factor associated to the modification of the 
theoretical prediction for the observable included in the fit due to the presence of a given New Physics mode. 
Hence, the procedure is completely general, in principle, and does not rely on any EFT expansion. 
\vspace*{3mm}

\paragraph{Code usage.} In terms of usage, a contaminated fit using the \simunet{} code should be based on an \nnpdf{} closure test runcard; in particular, the \texttt{closuretest} 
namespace must be specified. The required \simunet{} additions to such a runcard are twofold and described as follows. First, the \texttt{dataset\_inputs} 
must be modified with a new key, for example:
\begin{python}
dataset_inputs:
- {dataset: LHCB_Z_13TEV_DIELECTRON, frac: 0.75, cfac: ['QCD']}
- {dataset: CMSDY1D12, frac: 0.75, cfac: ['QCD', 'EWK'], contamination: 'EFT_LO'}
\end{python}
Here, two datasets are included: (i) the \texttt{LHCB\_Z\_13TEV\_DIELECTRON} dataset has no additional keys compared to a standard \nnpdf{} closure test 
runcard, and hence the level 1 pseudodata for this set is generated normally; (ii) the \texttt{CMSDY1D12} dataset has an additional key, \texttt{contamination}, 
set to the value \texttt{'EFT\_LO'}, which tells \simunet{} to base the level 1 pseudodata for this set on the predictions given by 
Eq.~\eqref{eq:simunet_contaminated}, using parameters drawn from the model \texttt{EFT\_LO} in the relevant \texttt{simu\_fac} file. The precise 
values of the true physical parameters used in Eq.~\eqref{eq:simunet_contaminated} are specified as part of the \texttt{closuretest} namespace, 
which is the second modification required to a standard closure test runcard:
\begin{python}
closuretest:
  ...
  contamination_parameters:
    - name: 'W'
      value:  -0.0012752
      linear_combination:
        'Olq3': -15.94

    - name: 'Y'
      value: 0.00015
      linear_combination:
        'Olq1': 1.51606
        'Oed': -6.0606
        'Oeu': 12.1394
        'Olu': 6.0606
        'Old': -3.0394
        'Oqe': 3.0394
\end{python}
In the example above we have specified that we inject a New Physics model that belongs to the so-called {\it universal theories} 
in which the parameters $\hat{W}$ and $\hat{Y}$, which are a combination of four-fermion operators defined for example in 
Refs.~\cite{Greljo:2021kvv,Iranipour:2022iak,Hammou:2023heg}, are set the values given in this example.
The precise definitions of the linear combinations in this example are given by:
\begin{eqnarray*}
  \hat{W}\,\mathcal{O}_W &=& (-0.0012752) \times \left(-15.94\, \mathcal{O}_{lq3}\right),\\
  \hat{Y}\,\mathcal{O}_Y &=& (0.00015) \times \left(1.51606\, \mathcal{O}_{lq1} -6.0606\,\mathcal{O}_{ed} 
  +12.1394\,\mathcal{O}_{eu}+6.0606\,\mathcal{O}_{lu}-3.0394\,\mathcal{O}_{ld}+3.0394\,\mathcal{O}_{qe}\right).
\end{eqnarray*}

\subsection{Closure tests with \simunet{}} 
\label{sec:ctall}
The `contamination' feature of the \simunet{} code allows the user not only to study whether a theory bias 
can be reabsorbed into the PDFs, but can be used in conjunction with the simultaneous fit feature to perform 
closure tests on the simultaneous fits of PDFs and SMEFT Wilson coefficients.
The closure test validates how much \simunet{} is able to replicate, not only fixed Wilson coefficients, but
also a known underlying PDF. In the context of a simultaneous fitting methodology a `level 2' closure test amounts 
to generating SM predictions using a known PDF set (henceforth referred to as the
underlying law) and mapping these to SMEFT observables by multiplying the SM theory
predictions with SMEFT $K$-factors scaled by a previously determined choice of Wilson coefficients. 
These SMEFT observables, generated by the underlying law, replace the usual
MC pseudodata replicas, and are used to train the neural network in the usual way.

Through this approach we are not only able to assess the degree to which the parametrisation 
is able to capture an underlying choice of Wilson coefficients, but also ensure it is
sufficiently flexible to adequately replicate a known PDF. 

\paragraph{Closure test settings.} 
In the rest of the section, we present the results of a simultaneous closure test the validates the \simunet{} methodology in
its ability to produce an underlying law comprising both PDFs and Wilson coefficients. In this example we set:
\begin{align*}
  \omega_{\rm true} &= {\tt NNPDF40\_nnlo\_as\_0118\_1000},\\
  \vec{c}_{\rm true} &= \begin{pmatrix} c_{tG} \\ c_{lq}^{(3)} \end{pmatrix} = \begin{pmatrix} 1 \\ 0.08 \end{pmatrix}.
\end{align*}
That is, \nnpdf{} serves as the true underlying PDF law, and all Wilson coefficients are set to zero except $c_{tG}$ and $c_{lq}^{(3)}$.
Note that the {\tt true} values of the Wilson coefficients that we choose in this test are 
outside the 95\% C.L. bounds on the Wilson coefficients that were found in previous 
analyses~\cite{Greljo:2021kvv,Kassabov:2023hbm} corresponding to $c_{tG}\in (-0.18,0.17)$ and 
$c_{lq}^{(3)}\in (-0.07,0.02)$ respectively, hence the success of the 
closure test is non-trivial. 

\begin{figure}[htb!]
  \centering
  \includegraphics[width=0.49\textwidth]{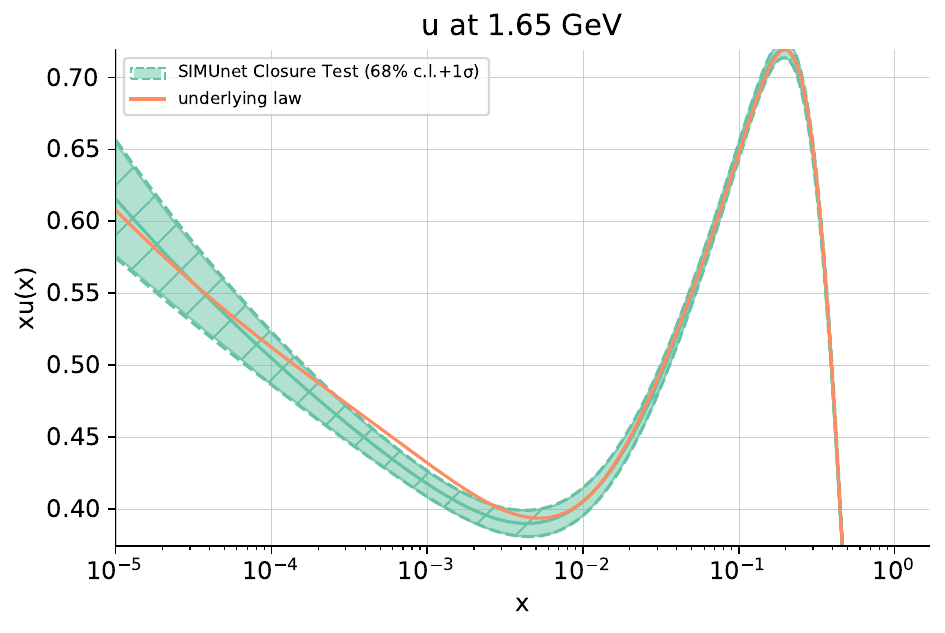}
  \includegraphics[width=0.49\textwidth]{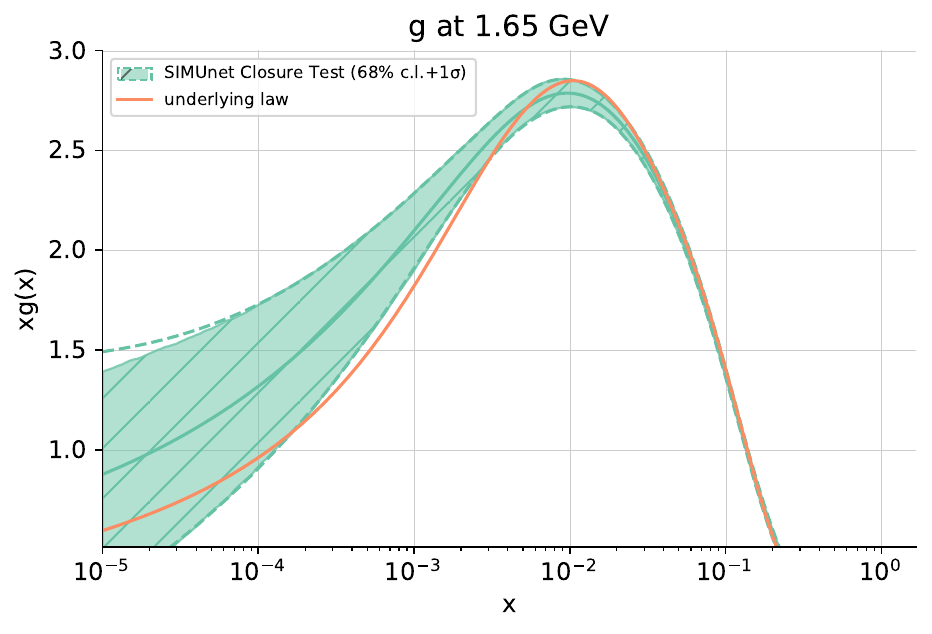}
\caption{
  The gluon (left) and up quark (right) PDFs obtained from the closure test 
framework in which both the underlying PDF set and Wilson coefficients are known. 
Shown in orange is the PDF replica used as the underlying law which generates the fake data used to train our model. 
The resulting PDFs are shown in green along with their 68\% confidence level bands. 
The fake data generated by the underlying law is subsequently modified so as to encode the ($c_{tG}$, $c_{lq}^{(3)}$) = (1, 0.08) condition.
}
\label{fig:pdf_closuretest}
\end{figure}
The results of the closure test for the gluon and up quark PDFs at the parametrisation scale 
are presented in Fig. (\ref{fig:pdf_closuretest}). Here we can see that the combination layer is
capturing the data’s dependence on the Wilson coefficients, whilst the complementary PDF
sector of the network architecture captures the data’s dependence on the underlying PDF. 
The combination layer, in effect, subtracts off the EFT
dependence, leaving behind the pure SM contribution for the PDF sector to parameterise.
\begin{figure}[htb!]
    \centering
    \includegraphics[width=0.5\textwidth]{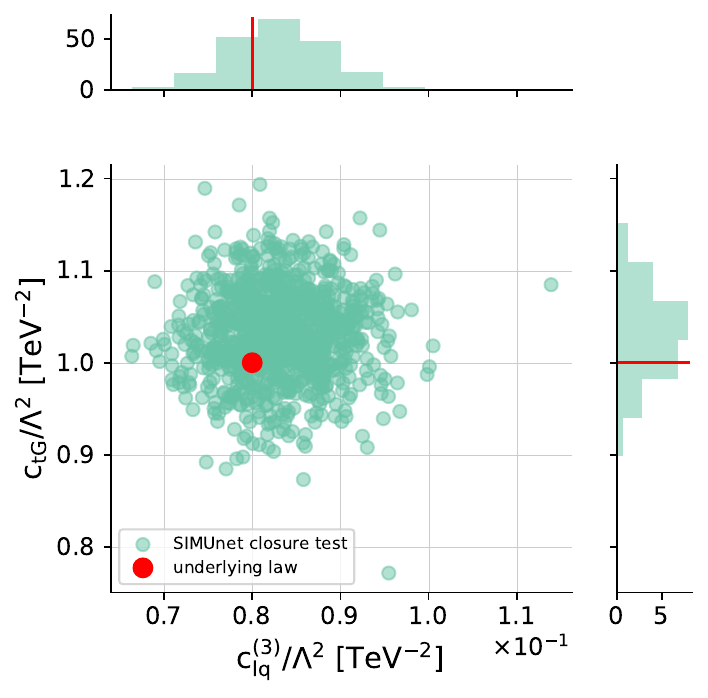}
\caption{
    Result of the closure testing framework for our methodology. 
    The distribution of ($c_{tG}$,$c_{lq}^{(3)}$) when fitting to data that has been modified 
    by setting ($c_{tG}$,$c_{lq}^{(3)}$) = ($1$, $0.08$). 
    The upper and right panels show the histograms for the distribution of the best fit values in their respective directions.
    }
\label{fig:wc_closuretest}
\end{figure}
The corresponding results for the Wilson coefficients $c_{tG}$ and $c_{lq}^{(3)}$ are 
displayed in Fig. (\ref{fig:wc_closuretest}). 
Both figures demonstrate that the parametrisation successfully captures the underlying 
law for both the PDFs and Wilson coefficients.
To verify the robustness of these findings and rule out random fluctuations, we conducted 
the closure test 25 times, 
each time using different level 1 pseudodata; the results consistently remained 
stable and aligned with those presented above.

\section{A first global simultaneous fit of SMEFT WCs and PDFs}
\label{sec:data}

In this section we present the first global simultaneous fit of SMEFT WCs and PDFs, including Deep-Inelastic-Scattering, 
Drell-Yan, top, jets, Higgs, diboson and EW precision observables.  This analysis can be now performed thanks to the 
\simunet{} methodology that we have described so far. 
We start in Sect.~\ref{subsec:data} by listing the data that have been included, and giving details of their implementation. 
In Sect.~\ref{subsec:fixedpdf} we 
present the result of the fixed-PDF analysis, 
producing a global SMEFT fit of a large number of datapoints from the Higgs, top, and EW sectors ($n_{\rm data}\,=\,366$)
to a set of $n_{\rm op}\,=\,40$ SMEFT operators.
Finally in Sect.~\ref{subsec:simu} we present the 
simultaneous PDF and SMEFT global fit ($n_{\rm data}\,=\,4985$).
We compare the resulting PDFs and SMEFT bounds to the SM PDFs and to the SMEFT bounds obtained in the fixed-PDF fit respectively, hence 
assessing the effect of the interplay between PDFs and SMEFT effects on both our description of the proton and 
on New Physics bounds. 

\subsection{Experimental data}
\label{subsec:data}
The dataset explored in this study builds on those already implemented in the high-mass Drell-Yan analysis 
presented in Refs.~\cite{Greljo:2021kvv,Iranipour:2022iak} and on the
top quark analysis presented in Ref.~\cite{Kassabov:2023hbm}. 
These studies extended the datasets included in the global \nnpdf{}~\cite{NNPDF:2021njg} analysis, 
by adding observables that enhance the sensitivity to the SMEFT and that constrain PDFs in the Drell-Yan and in the 
top sector respectively. 

The \nnpdf{} NNLO analysis~\cite{NNPDF:2021njg} included $n_{\rm data}\,=\,4618$ data points corresponding to a wide variety of processes in
deep-inelastic lepton-proton scattering (from the HERA $ep$ collider and from fixed-target experiments such as NMC, SLAC, 
BCDMS, CHORUS and NuTeV), fixed-target DY data from the
Fermilab E605 and E866 experiments, hadronic proton-antiproton collisions from the Tevatron, and proton-proton collisions from LHC. 
The LHC data in turn include inclusive gauge boson production data; $Z$- and $W$-boson transverse momentum production data, 
single-inclusive jet and di-jets production data, as well as gauge boson with jets and inclusive isolated photon
production data and top data. 
In Refs.~\cite{Greljo:2021kvv,Iranipour:2022iak} the high-mass Drell-Yan sector of the \nnpdf{} analysis 
was augmented by two extra measurements taken by CMS at $\sqrt{s}=8,13$ TeV, for a total of 65 extra datapoints. 
In Ref.~\cite{Kassabov:2023hbm}, besides inclusive and differential $t\bar{t}$ cross sections and $t$-channel 
single top production already implemented in \nnpdf{}, more observables were included, such as $t\bar{t}$ production asymmetries, 
$W$-helicity fractions, associated top pair production with vector bosons and heavy quarks, including 
$\bar{t}t Z$, $\bar{t}t W$, $\bar{t}t \gamma$, $\bar{t}t\bar{t}t$, $\bar{t}t\bar{b}b$, $s$-channel 
single top production, and associated single top and vector boson production, bringing the total number of datapoints to $n_{\rm data}\,=\,4710$. 

In this present study, we further extend the datasets by probing electroweak precision observables (EWPOs), 
the Higgs sector and the diboson sector. 
The datasets are described below and details such as the centre-of-mass energy, the observable, the integrated luminosity, the number 
of data points, the dataset name as implemented in \simunet \ and the source are given 
in Tables~\ref{tab:dataset_EWPO},~\ref{tab:dataset_Higgs} and~\ref{tab:dataset_diboson} respectively. 

We remind the reader that, thanks to the functionality of \simunet{} that enables users to include PDF-independent observables 
as well as to freeze PDFs for the observables in which the PDF dependence is mild, we can add observables such as the measurement of 
$\alpha_e$ that are completely independent of PDFs (but which are affected by SMEFT corrections) or Higgs signal strengths and Simplified Template Cross Section (STXS) measurements that 
are mildly dependent on PDFs but are strongly dependent on the relevant SMEFT-induced corrections. 

To summarise, in this analysis we include a total of {\bf 4985 datapoints}, of which 366 are affected by SMEFT corrections. 
Among those 366 datapoints, 210 datapoints are PDF independent. 
Hence our pure PDF analysis performed with \simunet{} includes $(4985-210) = 4775$ datapoints and is equivalent to the 
\nnpdf{} global analysis augmented by 78 datapoints from the top sector~\cite{Kassabov:2023hbm} and 
by 65 datapoints measuring the high-mass Drell-Yan tails~\cite{Greljo:2021kvv,Iranipour:2022iak}. Our pure 
SMEFT-only analysis performed with \simunet{} includes 366 datapoints, while our simultaneous PDF-SMEFT fit includes 
4985 datapoints. 

\vspace*{0.4cm}
\noindent {\bf EWPOs:}
In Table~\ref{tab:dataset_EWPO}, we list EWPOs included in this study. 
The dataset includes the pseudo-observables
measured on the $Z$ resonance by LEP~\cite{ALEPH:2005ab}, 
which include cross sections, forward-backward and polarised asymmetries. 
Branching ratios of the decay of the $W$ boson into leptons~\cite{ALEPH:2013dgf} are also included, 
along with the LEP measurement of the Bhabha scattering cross section~\cite{ALEPH:2013dgf}. 
Finally we include the measurement of the effective electromagnetic coupling constant~\cite{Workman:2022ynf}, 
for a total of {\bf 44 datapoints}.  These datasets and their SMEFT predictions
are taken from the public \smefit{} code \cite{Giani:2023gfq}.  These datapoints are all PDF independent, hence they directly 
affect only the SMEFT part of the fits.
\begin{table}[htb!]
    \begin{center}
  {\fontsize{8pt}{8pt}\selectfont
  %\begin{table}[t]
    \centering
     \renewcommand{\arraystretch}{2}
     \setlength{\tabcolsep}{5pt}
     \begin{tabular}{lcccccc}
       \toprule \textbf{Exp.}   & $\bf{\sqrt{s}}$ \textbf{(TeV)}
      &  \textbf{Observable} & $\mathcal{L}$ (fb${}^{-1}$) & $\mathbf{n_{\rm dat}}$ & \textbf{Dataset name}
       &\textbf{Ref.}\\
      \toprule
        \bf{LEP}
        & 0.250
        & Z observables
        & 
        & 19
        & LEP\_ZDATA
        & \cite{ALEPH:2005ab}\\
      \midrule
        \bf{LEP}
        & 0.196
        & $\mathcal{B}(W \rightarrow l^{-} \bar{v}_l)$
        & 3
        & 3
        & LEP\_BRW
        & \cite{ALEPH:2013dgf}\\
      \midrule
        \bf{LEP}
        & 0.189
        & $\sigma(e^+ e^- \rightarrow e^+ e^-)$
        & 3
        & 21
        & LEP\_BHABHA
        & \cite{ALEPH:2013dgf}\\
      \midrule
        \bf{LEP}
        & 0.209
        & $\hat{\alpha}^{(5)}_{\rm}(M_Z)$
        & 3
        & 1
        & LEP\_ALPHAEW
        & \cite{Workman:2022ynf}\\
  
  \bottomrule
     \end{tabular}
     \vspace{0.3cm}
     \caption{Measurements of electroweak precision observables included in \simunet{}. The columns contain information on the 
     centre-of-mass energy, the observable, the integrated luminosity, the number 
     of data points, the dataset name as implemented in \simunet \ and the source.
      \label{tab:dataset_EWPO}
  }
  }
  \end{center}
  \end{table}

\vspace*{0.4cm}
\noindent {\bf Higgs sector:}
In Table~\ref{tab:dataset_Higgs}, we list the Higgs sector datasets included in this study. 
The Higgs dataset at the LHC includes the combination of Higgs signal strengths
by ATLAS and CMS for Run 1, and for Run 2 both signal strengths and STXS
measurements are used. ATLAS in particular provides the combination of stage 1.0 STXS bins
for the $4l$, $\gamma\gamma$, $WW$, $\tau\tau$ and $b\bar{b}$ decay channels, while for CMS we use the combination
of signal strengths of the $4l$,
$\gamma\gamma$, $WW$, $\tau^-\tau^+$, $b\bar{b}$ and $\mu^-\mu^+$ decay channels. 
We also include the $H \rightarrow Z\gamma$ 
and $H \rightarrow \mu^-\mu^+$ signal strengths from ATLAS, for a total of {\bf 73 datapoints}. 
The Run I and CMS datasets and their corresponding predictions are taken from the \smefit{} code \cite{Giani:2023gfq},
whereas the STXS observables and signal strength measurements of
$H \rightarrow Z\gamma$ and $H \rightarrow \mu^-\mu^+$ are taken from the \fitmaker ~\cite{Ellis:2020unq} code.
The signal strength's dependence on the PDFs is almost completely negligible, given that the 
PDF dependence cancels in the ratio, hence all the datapoints are treated as PDF independent and 
are directly affected only by the SMEFT Wilson coefficients.
\begin{table}[htb!]
    \begin{center}
  {\fontsize{8pt}{8pt}\selectfont
  %\begin{table}[t]
    \centering
     \renewcommand{\arraystretch}{2}
     \setlength{\tabcolsep}{5pt}
     \begin{tabular}{lcccccc}
       \toprule \textbf{Exp.}   & $\bf{\sqrt{s}}$ \textbf{(TeV)}
      &  \textbf{Observable} & $\mathcal{L}$ (fb${}^{-1}$) & $\mathbf{n_{\rm dat}}$ & \textbf{Dataset name}
       &\textbf{Ref.}\\
      \toprule
        \bf{ATLAS and CMS}
        & 7 and 8
        & $\mu_{H \rightarrow \mu^+ \mu^-}$
        & 5 and 20
        & 22
        & ATLAS\_CMS\_SSinc\_RunI
        & \cite{ATLAS:2016neq}\\
      \midrule
        \bf{CMS}
        & 13
        & $\mu_{H}$
        & 35.9
        & 24
        & CMS\_SSINC\_RUNII
        & \cite{CMS:2018uag}\\
      \midrule
        \bf{ATLAS}
        & 13
        & $\mu_{H}$
        & 80
        & 25
        & ATLAS\_STXS\_RUNII
        & \cite{ATLAS:2019nkf}\\
      \midrule
        \bf{ATLAS}
        & 13
        & $\mu_{H \rightarrow Z \gamma}$
        & 139
        & 1
        & ATLAS\_SSINC\_RUNII\_ZGAM
        & \cite{ATLAS:2020qcv}\\
      \midrule
        \bf{ATLAS}
        & 13
        & $\mu_{H \rightarrow \mu^+ \mu^-}$
        & 139
        & 1
        & ATLAS\_SSINC\_RUNII\_MUMU
        & \cite{ATLAS:2020fzp}\\

  \bottomrule
     \end{tabular}
     \vspace{0.3cm}
     \caption{Same as~\ref{tab:dataset_EWPO} for the measurements of the Higgs sector.
      \label{tab:dataset_Higgs}  }
  }
  \end{center}
  \end{table}

\vspace*{0.4cm}
\noindent {\bf Diboson sector:}
In Table~\ref{tab:dataset_diboson}, we list the diboson sector datasets included in this study. 
We have implemented four measurements from LEP at low energy, three from ATLAS and one from CMS both at 13 TeV. 
The LEP measurements are of the differential 
cross-section of $e^+ e^- \rightarrow W^+ W^-$ as a function of the cosine of the $W$-boson polar angle. 
The ATLAS measurements are of 
the differential cross-section of $W^+ W^-$ as a function of the invariant mass of the electron-muon pair 
and of the transverse mass of the $W$-boson. 
The third ATLAS measurement is of the differential cross-section of $Zjj$ as a function of the azimuthal angle 
between the two jets, $\Delta \phi_{jj}$.  Although this is not a diboson 
observable, it is grouped here as this observable constrains operators typically 
constrained by the diboson sector,
such as the triple gauge interaction $O_{WWW}$. The CMS dataset is a measurement of the differential cross-section of $WZ$ as 
a function of the transverse momentum 
of the $Z$-boson, for a total of {\bf 82 datapoints}. 
Almost all datasets and corresponding predictions in this sector are taken from the \smefit{} code~\cite{Giani:2023gfq},
except for the ATLAS measurement of $Zjj$ production, taken from the \fitmaker ~\cite{Ellis:2020unq} code.
While the LEP measurements are PDF-independent by definition, 
the LHC measurements of diboson cross sections do depend on PDFs. However, their impact on the 
PDFs is extremely mild, due to much more competitive constraints in the same Bjorken-$x$ region given by other measurements
such as single boson production and single boson production in association with jets. 
Therefore we consider all diboson measurements 
to be PDF-independent, i.e. the PDFs are fixed and their parameters are not varied in the computation of these observables. 
\begin{table}[htb!]
    \begin{center}
  {\fontsize{8pt}{8pt}\selectfont
  %\begin{table}[t]
    \centering
     \renewcommand{\arraystretch}{2}
     \setlength{\tabcolsep}{5pt}
     \begin{tabular}{lcccccc}
       \toprule \textbf{Exp.}   & $\bf{\sqrt{s}}$ \textbf{(TeV)}
      &  \textbf{Observable} & $\mathcal{L}$ (fb${}^{-1}$) & $\mathbf{n_{\rm dat}}$ & \textbf{Dataset name}
       &\textbf{Ref.}\\
      \toprule
        \bf{LEP}
        & 0.182
        & $d \sigma _{WW} / d cos(\theta _W)$
        & 0.164
        & 10
        & LEP\_EEWW\_182GEV
        & \cite{ALEPH:2013dgf}\\
      \midrule
        \bf{LEP}
        & 0.189
        & $d \sigma _{WW} / d cos(\theta _W)$
        & 0.588
        & 10
        & LEP\_EEWW\_189GEV
        & \cite{ALEPH:2013dgf}\\
      \midrule
        \bf{LEP}
        & 0.198
        & $d \sigma _{WW} / d cos(\theta _W)$
        & 0.605
        & 10
        & LEP\_EEWW\_198GEV
        & \cite{ALEPH:2013dgf}\\
      \midrule
        \bf{LEP}
        & 0.206
        & $d \sigma _{WW} / d cos(\theta _W)$
        & 0.631
        & 10
        & LEP\_EEWW\_206GEV
        & \cite{ALEPH:2013dgf}\\
      \midrule
        \bf{ATLAS}
        & 13
        & $d \sigma _{W^+W^-}/d m_{e \mu}$
        & 36.1
        & 13
        & ATLAS\_WW\_13TeV\_2016\_MEMU
        & \cite{ATLAS:2019rob}\\
      \midrule
        \bf{ATLAS}
        & 13
        & $d \sigma _{WZ} / d m_{T}$
        & 36.1
        & 6
        & ATLAS\_WZ\_13TeV\_2016\_MTWZ
        & \cite{ATLAS:2019bsc}\\
      \midrule
        \bf{ATLAS}
        & 13
        & $d \sigma(Zjj)/d \Delta \phi_{jj}$
        & 139
        & 12
        & ATLAS\_Zjj\_13TeV\_2016
        & \cite{ATLAS:2020nzk}\\
      \midrule
        \bf{CMS}
        & 13
        & $d \sigma _{WZ} / d p_{T}$
        & 35.9
        & 11
        & CMS\_WZ\_13TeV\_2016\_PTZ
        & \cite{ATLAS:2020nzk}\\

  \bottomrule
     \end{tabular}
     \vspace{0.3cm}
     \caption{Same as~\ref{tab:dataset_EWPO} for the measurements of the diboson sector.
      \label{tab:dataset_diboson}
  }
  }
  \end{center}
  \end{table}

\vspace*{0.4cm}
\noindent {\bf SMEFT Wilson coefficients:}
We include a total of 40 operators from the dimension-6 SMEFT in our global analysis 
of the measurements of the
Higgs, top, diboson and electroweak precision observables outlined above.
\begin{figure}[thb!]
    \centering
    \includegraphics[width=0.7\textwidth]{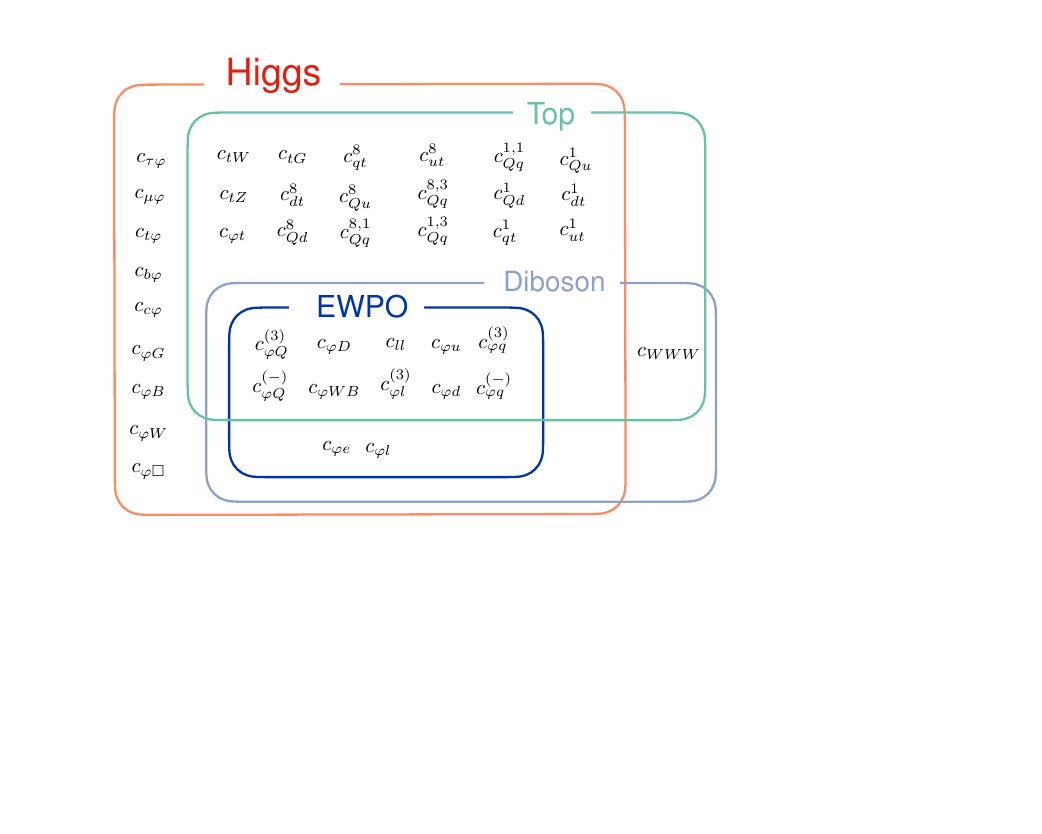}
    \caption{Schematic representation of the data categories included in this analysis and of 
    their overlapping dependences on the 40 dim-6 SMEFT operators included in our global analysis, 
    adapted from Ref.~\cite{Ellis:2020unq}.}
\label{fig:operators}
\end{figure}
In Fig.~\ref{fig:operators} we display the SMEFT Wilson coefficients included in our analysis, following
the operator conventions of Refs.~\cite{Kassabov:2023hbm,Ethier:2021bye}. 
The schematic diagram, adapted from Ref.~\cite{Ellis:2020unq}, demonstrates the sectors of data affected by each WC and highlights
the interconnected nature of the global SMEFT determination.
The overlaps between the different rectangles show explicitly how a given operator contributes to several data categories.
For example, $O_{WWW}$, contributes to both the diboson sector and the top sector (through its contribution to $tZq$ production),
while the operators $O_{\varphi e}$,$O_{\varphi l}$ contribute to the Higgs, diboson 
and electroweak sectors, but have no effect on the top sector.

%%%%%%%%%%%%%%%%%%%%%%%%%%%%%%%%%%%%%%%%%%%%%%%%%%%%%%%%%%%
\subsection{Global SMEFT fit}
\label{subsec:fixedpdf}
%%%%%%%%%%%%%%%%%%%%%%%%%%%%%%%%%%%%%%%%%%%%%%%%%%%%%%%%%%%

We present the results of this new global fit in two parts, beginning with a discussion 
of the result of the fixed PDF fit, that is a global fit purely of the SMEFT Wilson 
coefficients. 
In Sect.~\ref{subsec:simu} we will present the results of the simultaneous global 
fit of SMEFT WCs and will compare the results to those obtained here.

The global analysis is performed at linear order in the SMEFT operator
coefficients, accounting for the interference between the SM and the insertion of a dimension-6 SMEFT operator. 
The linear constraints can be viewed as 
provisional for operators where quadratic contributions are non-negligible. Nevertheless,
keeping those operators in the global fit typically yields conservative marginalised limits
that allows one to assess the impact on other operators and whether this is significant to a first
approximation. An analysis including quadratic constributions requires a new methodology that does not 
rely on the Monte-Carlo sampling method to propagate uncertainties~\cite{progr}, given that the latter fails at
reproducing the Bayesian confidence intervals once quadratic corrections are dominant,  
as described in App.~E of Ref.~\cite{Kassabov:2023hbm}.

Our analysis comprises  $n_{\rm op} = 40$ operators in a simultaneous combination of the constraints 
from $n_{\rm data}\,=\,366$ datapoints 
from the Higgs, EWPO, diboson and top sectors. Note that in the \simunet{} code there 
are more SMEFT operators implemented, in particular the set of four-fermion operators that 
affect Drell-Yan and DIS data. As demonstrated in~\cite{Carrazza:2019sec,Greljo:2021kvv}, 
the effect of the four 
fermion operators in DIS and on the current Drell-Yan data, including the high-mass Drell-Yan 
data included in \nnpdf{} is not significant enough for this data to provide strong constraints 
on the relevant set of four-fermion operators. 
On the contrary, the HL-LHC projections, which include both neutral and charged current 
Drell-Yan data have a strong potential in constraining such operators~\cite{Iranipour:2022iak}. 
As a results, Drell-Yan data are not affected by SMEFT corrections in this current analysis, 
but we leave the user the opportunity to include such effects straightforwardly, 
should the measurements by the LHC increase the constraining potential, or should the user combine 
those with flavour observables, as is done in Ref.~\cite{Bartocci:2023nvp}.

In the following, we also omit from the results the $4$ heavy quark production datasets (four tops and $t\bar{t}b\bar{b}$). 
Their effect is practically negligible on all the operators affecting the other sectors and they would simply constrain $2$ 
directions in the $5$-dimensional space of the four heavy fermion operators. These observables are therefore {\it de facto} decoupled 
from the rest of the dataset.

The input PDF set, which is kept fixed during the pure SMEFT fit, is the {\tt nnpdf40nnlo\_notop} 
baseline, corresponding to the \nnpdf{} fit without the top data, to avoid possible contamination 
between PDF and EFT effects in the top sector~\cite{Kassabov:2023hbm}. 

The sensitivity to the EFT operators of the various processes entering the fit can
be evaluated by means of the Fisher information, which, for $N$ EFT coefficients, is an $N \times N$ matrix given by
\begin{equation}
  \label{eq:fisher}
  F_{ij} = {L^{i}}^T (\text{cov}_{\text{exp}})^{-1} L^{j},
\end{equation}
where $L^{i}$ is the vector of the linear contributions to the observables of the $i$-th SMEFT Wilson coefficient, and
$\text{cov}_{\text{exp}}$ is the experimental covariance matrix. The covariance matrix in the space of SMEFT coefficients $C_{ij}$, 
by the Cramér-Rao bound, satisfies $C_{ij} \geq (F^{-1})_{ij}$, hence the larger the entries of the Fisher information matrix, 
the smaller the possible uncertainties of the SMEFT coefficients and, therefore, the bigger the constraining power of the relevant dataset. 
\begin{figure}[tb!]
  \centering
  \includegraphics[width=0.9\textwidth]{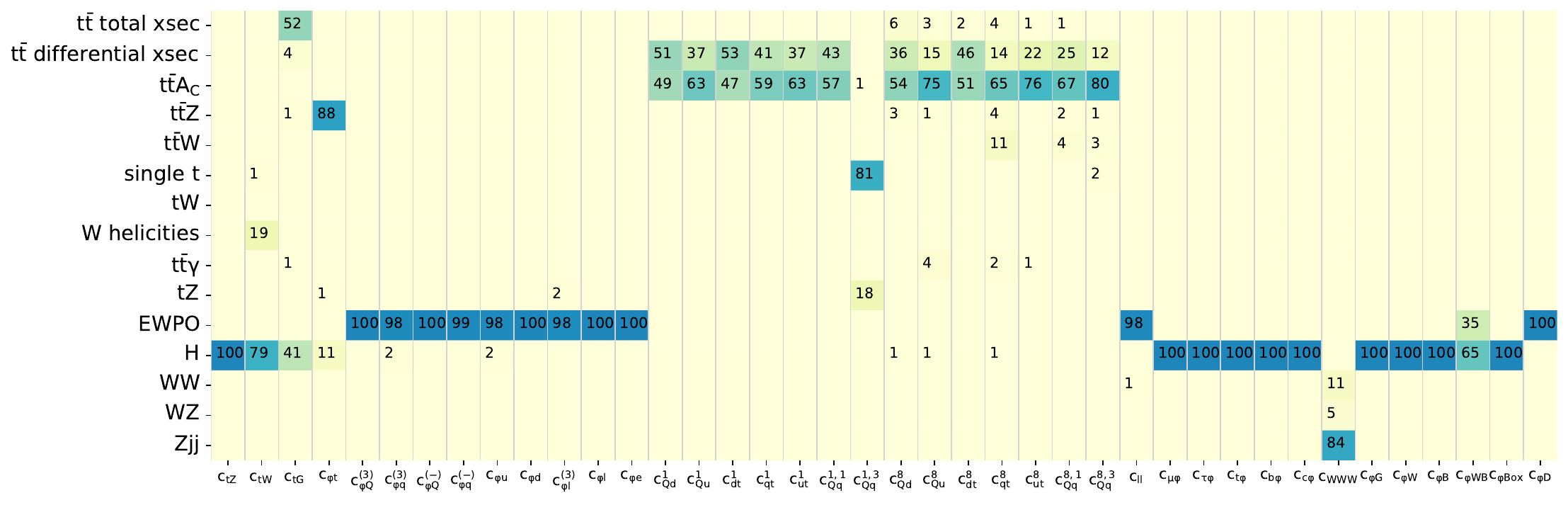}
  \caption{
Relative constraining power (in \%) on each of the operators for each of the processes entering the fit, 
as defined in Eq.~\eqref{eq:relativeconstrainingpower}.}
\label{fig:fisher}
\end{figure}
In Fig.~\ref{fig:fisher} we use the Fisher information to assess the relative impact of each sector of 
data included in our analysis on the SMEFT parameter space by plotting the relative percentage constraining power of the dataset $D$
via:
\begin{equation}
\label{eq:relativeconstrainingpower}
	\textrm{relative constraining power of }D\textrm{ on operator }c_i = F_{ii}(D) \bigg/ \displaystyle \sum_{\text{sectors }D'} F_{ii}(D'),
\end{equation}
which gives a general qualitative picture of some of the expected behaviour in the global fit. 
In Fig.~\ref{fig:fisher} we observe that the coefficients $c_{tZ}$, $c_{tW}$ are dominantly constrained by the Higgs sector, while
the coefficient $c_{tG}$ is now constrained both by the $t \bar{t}$ total cross sections and by the 
Higgs measurements, as expected.
Higgs measurements also provide the dominant constraints on the bosonic coefficients
$c_{\varphi G}$, $c_{\varphi W}$, $c_{\varphi B}$, $c_{\varphi \Box}$
and on the Yukawa operators.
Coefficients coupling the Higgs to fermions,
$c_{\varphi Q}^{(3)}$,
$c_{\varphi q}^{(3)}$, $c_{\varphi Q}^{(-)}$, $c_{\varphi q}^{(-)}$, $c_{\varphi u}$, $c_{\varphi d}$,
$c_{\varphi l}^{(3)}$ and $c_{\varphi e}$, receive their dominant constraining power from the electroweak precision observables,
as does the four-fermion coefficient $c_{ll}$.
The coefficient $c_{WWW}$ is constrained by the diboson sector, with measurements of $Zjj$ production
providing the leading constraints at linear order in the SMEFT.
From the top sector of the SMEFT, the four-fermion operators are constrained by measurements of top quark pair
production total cross sections, differential distributions and charge asymmetries.  The exception to this is $c_{Qq}^{(1,3)}$,
which is constrained by both single top production and single top production in association with a $Z$ boson.
Overall we find qualitatively similar results to those shown in Refs.~\cite{Ellis:2020unq,Ethier:2021bye}.

\begin{figure}[tb!]
    \centering
    \includegraphics[width=0.8\textwidth]{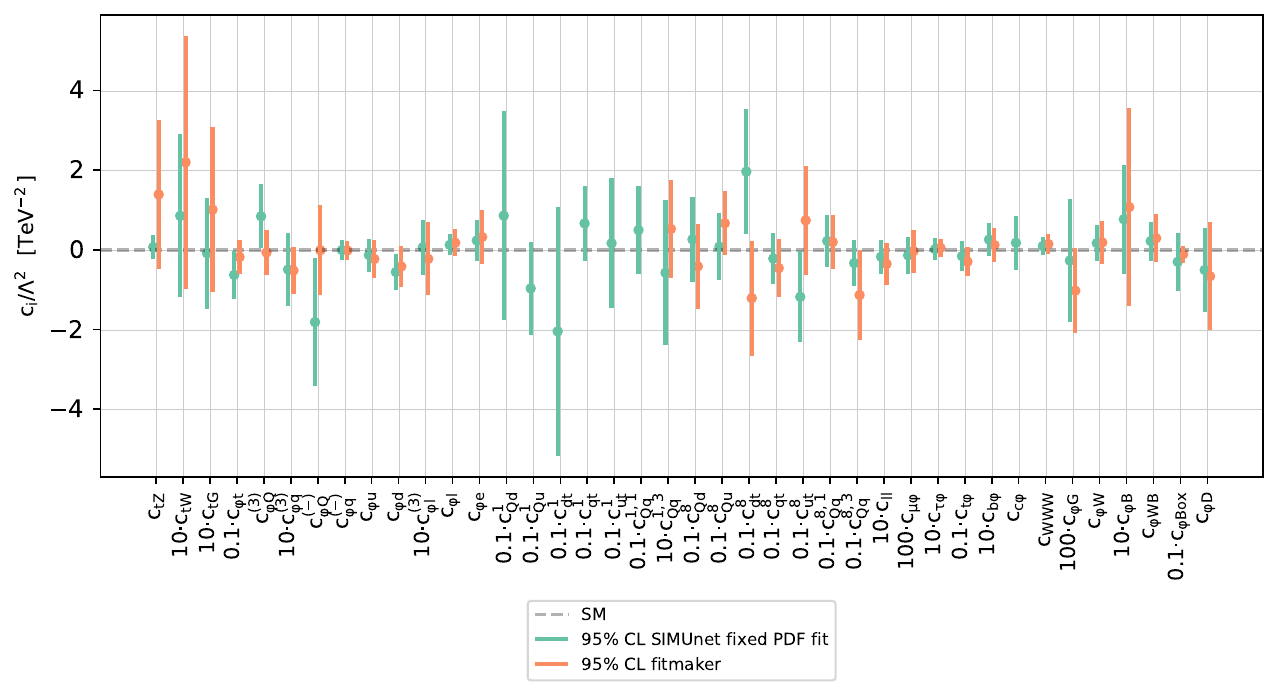}
    \caption{
  Comparison of the bounds for the SMEFT Wilson coefficients determined in the pure  
  SMEFT analysis presented in this work and the \fitmaker{} analysis~\cite{Ellis:2020unq}. 
  The dashed horizontal line is the SM prediction. In each bound, the dot represents 
  the best-fit value and the 95\% CL interval is constructed from the envelope of 
  two standard deviations away from this best-fit value.}
\label{fig:simunet_vs_fitmaker}
\end{figure}

To assess the strength of the bounds on the Wilson coefficients that we find in this work and 
for reference, in Fig.~\ref{fig:simunet_vs_fitmaker} we compare the bounds that we obtain here 
against the fixed-PDF SMEFT analysis presented by the \fitmaker{} 
collaboration~\cite{Ellis:2020unq}, which is the global public SMEFT analysis that currently 
includes all sectors that are considered here.
We observe that the bounds are comparable, with a few differences. First of all \fitmaker{} 
uses LO QCD predictions for the SMEFT corrections in the top sector, hence most of the singlet four 
fermion operators in the top sector are not included, while \simunet{} uses NLO QCD predictions 
and obtains bounds on those, although weak~\cite{Degrande:2020evl}. Moreover, while the 
other bounds overlap and are of comparable size, the ${\cal O}_{tZ}$ is much more constrained in 
the \simunet{} analysis, as compared to the \fitmaker{} one. 
This is due to the combination of all operators and observables, which collectively improves the constraints 
on this specific operator, thanks to the interplay between $O_{tZ}$ and the other operators that enter the same 
observables.

\begin{figure}[b!]
  \centering
  \includegraphics[width=0.8\textwidth]{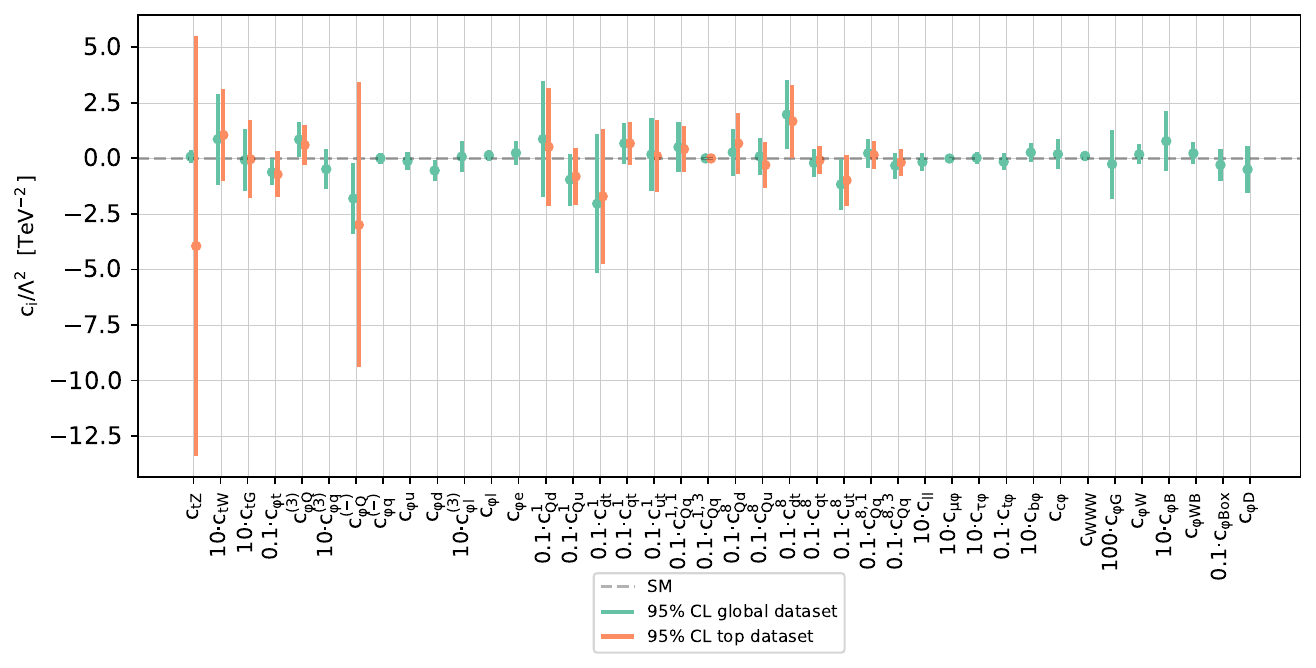}
  \caption{Same as Fig.~\ref{fig:simunet_vs_fitmaker}, now comparing the 95\% CL bounds obtained in the SMEFT top analysis 
  of Ref.~\cite{Kassabov:2023hbm} (old dataset) with the 95\% CL bounds obtained in this global SMEFT analysis.}
\label{fig:smeft_only_top_vs_global}
\end{figure}

In Fig.~\ref{fig:smeft_only_top_vs_global} we show the comparison between the results of this current global SMEFT analysis against 
the SMEFT analysis of the top sector presented in~\cite{Kassabov:2023hbm}.
In general, the results are very compatible, with bounds on most Wilson coefficients comparable between the two fits. 
However, the increased dataset size results in a marked improvement on the constraints of several Wilson coefficients, 
in particular $c_{\phi t}$, $c_{tZ}$ and $c_{\phi Q}^{(-)}$; see Table~\ref{tab:global_vs_top} for comparisons. 
\begin{table}[ht]
  \begin{center}
{\fontsize{8pt}{8pt}\selectfont
%\begin{table}[t]
  \centering
   \renewcommand{\arraystretch}{2}
   \setlength{\tabcolsep}{5pt}
   \begin{tabular}{lcc}
   \toprule
    \textbf{Operator} & \textbf{Global fit} & \textbf{Top fit} \\% & \textbf{Broadening} \\
  \toprule
  $c_{\phi t}$ & (-13, -0.22)  & (-18, 3.1) \\%& $-39.4$\%  \\
  \midrule
  $c_{tZ}$ & (-0.18, 0.37)  & (-13, 5.5) \\%& $-97.0$\%  \\
  \midrule
  $c_{\phi Q}^{(-)}$ & (-3.6, -0.042)  & (-9.4, 3.4)\\% & $-72.2$\%  \\
   \bottomrule
  \end{tabular}
  }
\end{center}
    \caption{\label{tab:global_vs_top} The 95\% confidence intervals on three key Wilson coefficients included in this analysis.}
  \end{table}
We observe that the bounds on $c_{\phi t}$ become more stringent by nearly a factor of 2, due to the extra constraints that come 
from the Higgs sector. Analogously, the Higgs constrains reduce the size of the bounds on $c_{\phi Q}^{(-)}$ by a factor of 3. 
The most remarkable effect is seen in $c_{tZ}$ which is now strongly constrained by the top loop contribution to the $H\to\gamma\gamma$ 
decay, for which we include experimental information on the signal strength from LHC Run II.

\begin{figure}[tb!]
  \centering
  \includegraphics[width=0.47\textwidth]{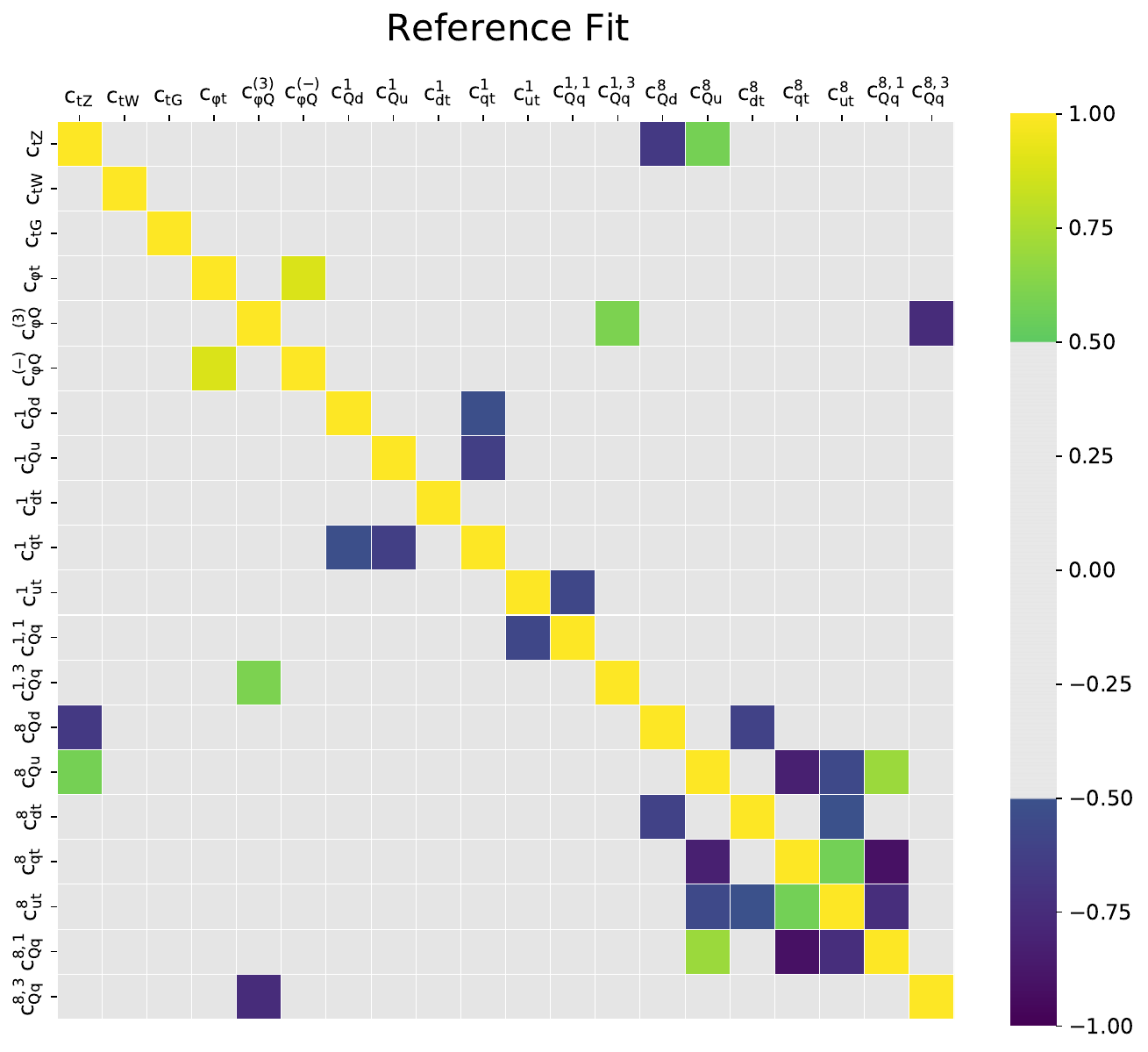}
  \includegraphics[width=0.47\textwidth]{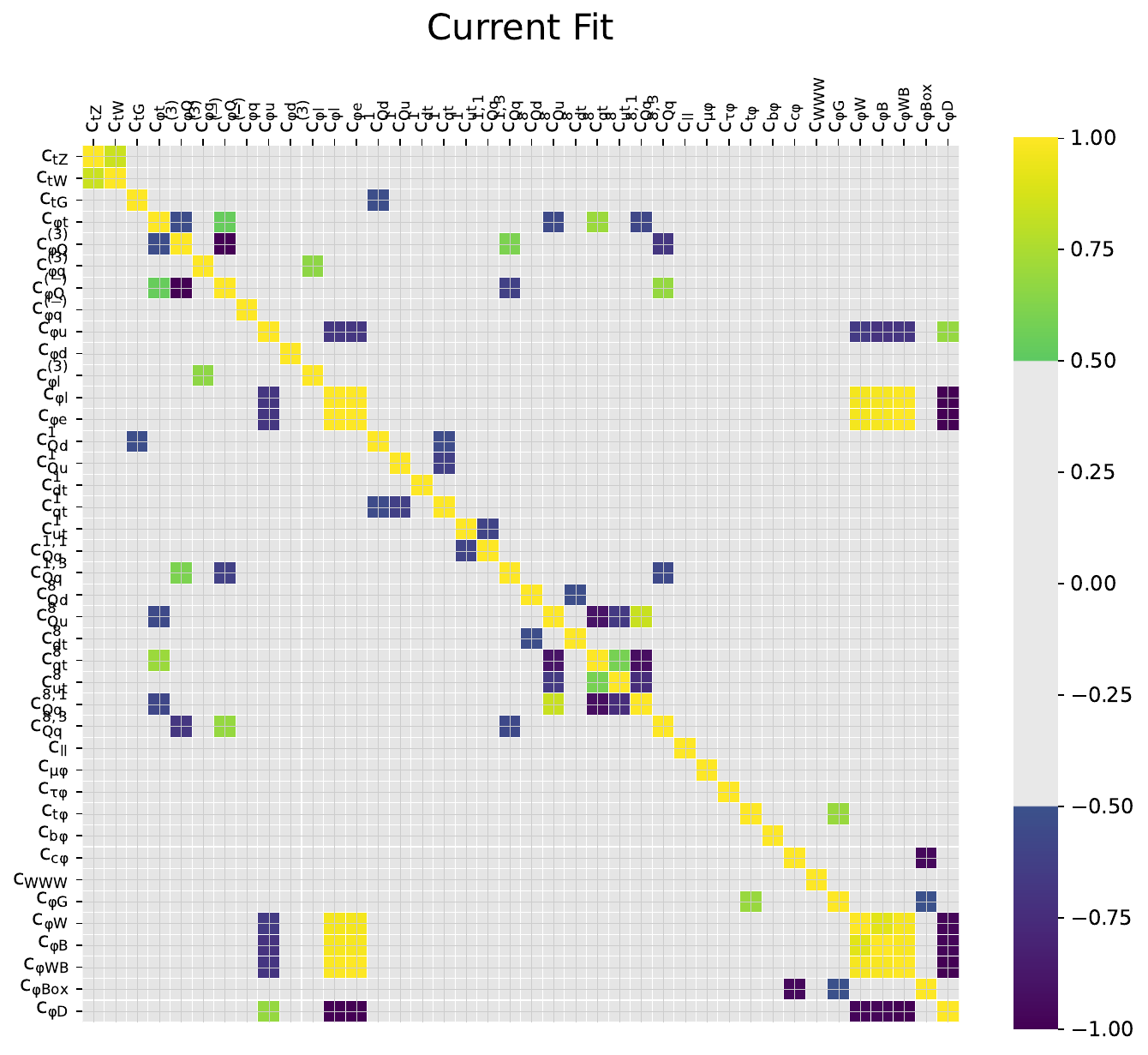}
  \caption{The values of the correlation coefficients evaluated between all pairs of Wilson coefficients entering the
  SMEFT fit of the top-only analysis (left) and the global analysis (right). }
\label{fig:fisher_info}
\end{figure}
It is interesting to compare the correlations between Wilson coefficients evaluated in this analysis
to the top-only SMEFT analysis of Ref.~\cite{Kassabov:2023hbm}. In Fig.~\ref{fig:fisher_info} we 
note that the additional sectors, particularly the Higgs one, 
introduce a degree of anti-correlation between $c_{\phi Q}^{(3)}$ and $c_{\phi t}$ and between $c_{Q q}^{(8,3)}$ and $c_{Q q}^{(1,3)}$, 
while the one between $c_{u t}^{8}$ and $c_{d t}^{8}$ goes away in the global analysis. In the EW sector, we find very strong correlations, 
in agreement with similar studies in the literature. This suggests that there is room for improvement in the sensitivity to the operators 
affecting EW observables once more optimal and targeted measurements are employed in future searches~\cite{GomezAmbrosio:2022mpm,Long:2023mrj}. 

Finally we display the correlations  observed between the PDFs and Wilson coefficients.
The PDF-EFT correlation coefficient for a Wilson coefficient $c$ and a
PDF $f(x, Q)$ at a given $x$ and $Q^2$ is defined as
\begin{equation}
\label{eq:correlationL2CT}
\rho\lp c, f(x,Q^2)\rp=\frac{\la c^{(k)}f^{(k)}(x,Q^2)\ra_k - \la c^{(k)}\ra_k \la f^{(k)}(x,Q^2)\ra_k
}{\sqrt{\la (c^{(k)})^2\ra_k - \la c^{(k)}\ra_k^2}\sqrt{\la\left( f^{(k)}(x,Q^2)\right)^2\ra_k - \la f^{(k)}(x,Q^2)\ra_k^2}}  \, ,
\end{equation}
where $c^{(k)}$ is the best-fit value of the Wilson coefficient for
the $k$-th replica and $f^{(k)}$ is the $k$-th PDF replica computed at
a given $x$ and $Q$, and $\la \cdot \ra_k$ represents the average
over all replicas. We compute the correlation between a SM PDF and the Wilson coefficients of
this new global analysis. By doing so we have a handle on 
which PDF flavours and kinematical regions are strongly impacted by the presence of
a given SMEFT-induced correction to the SM theoretical predictions, thus exhibiting a potential for interplay
in a simultaneous EFT-PDF determination.
\begin{figure}[hbt!]
  \centering
          \includegraphics[width=0.48\textwidth]{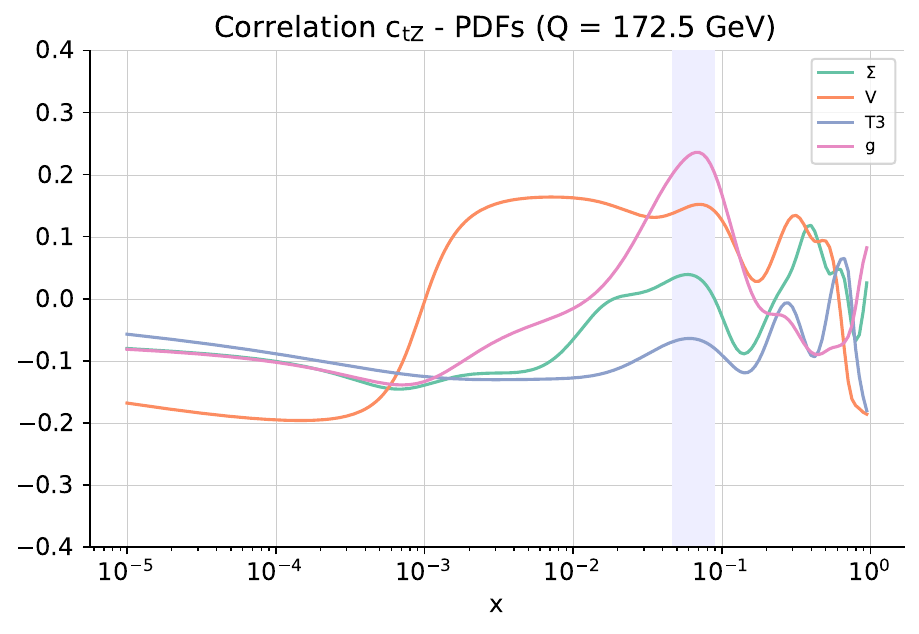}
          \includegraphics[width=0.48\textwidth]{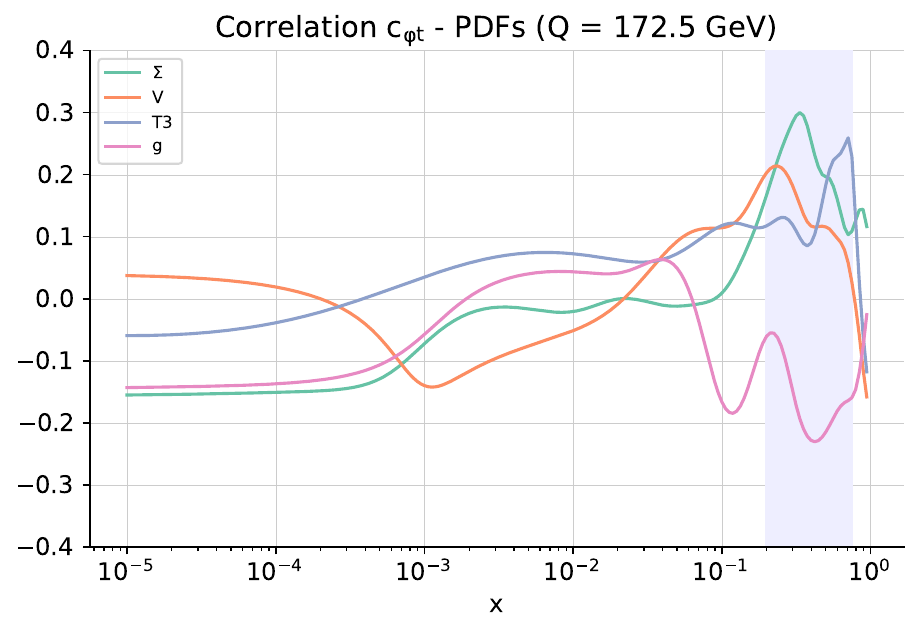}
    \caption{The value of the correlation coefficient $\rho$
            between the PDFs and selected EFT coefficients as a function of $x$
            and $Q=172.5$ GeV.
            We show the results for the gluon, the total singlet $\Sigma$, total valence $V$, and non-singlet
            triplet $T_3$ PDFs.
            We provide results for representative EFT coefficients,
            namely $c_{tZ}$ and  $c_{\phi t}$.
          }
  \label{fig:pdfbsmcorr}
  \end{figure}
Fig.~\ref{fig:pdfbsmcorr} displays a selection of the largest correlations.
We observe that the gluon PDF in the medium to large-$x$ region is significantly correlated with the Wilson coefficients 
$c_{tZ}$, as one would expect given that $c_{tZ}$ is strongly constrained by the top loop contribution to the Higgs decay, which
in turn is predominantly correlated to the gluon-gluon parton luminosity. On the other hand $c_{\phi t}$ is anticorrelated
with the gluon and correlated to the singlet and triplet in the large-$x$ region.
This is not surprising, given the impact of top quark pair production total cross sections and differential distributions
in constraining these PDFs and Wilson coefficients.  
Whilst these correlations are computed from a determination of the SMEFT in which the PDFs are fixed to SM PDFs, 
the emergence of non-zero correlations provides an indication of the potential for interplay between the PDFs and the SMEFT
coefficients; this interplay will be investigated in a simultaneous determination
in the following section.

%%%%%%%%%%%%%%%%%%%%%%%%%%%%%%%%%%%%%%%%%%%%%%%%%%%%%%%%%%%
\subsection{Global simultaneous SMEFT and PDF fit}
\label{subsec:simu} 
%%%%%%%%%%%%%%%%%%%%%%%%%%%%%%%%%%%%%%%%%%%%%%%%%%%%%%%%%%%

In this section we present the results of the simultaneous global fit of SMEFT Wilson coefficients and PDFs. We 
compare the bounds on the coefficients obtained in the two analyses as well as the resulting PDF sets, 
to assess whether the inclusion of more PDF-independent observables modifies the interplay 
observed in the top-focussed analysis of Ref.~\cite{Kassabov:2023hbm}.

\begin{figure}[tb!]
    \centering
    \includegraphics[width=0.8\textwidth]{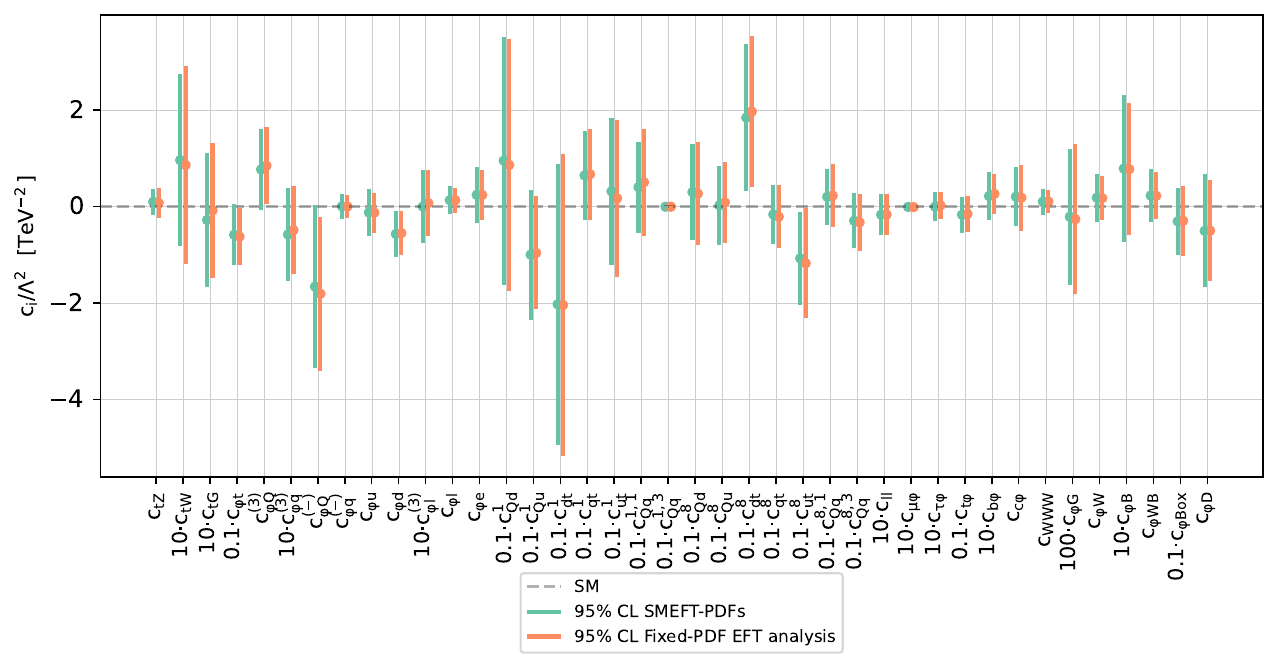}
\caption{
  Comparison of the 95\% CL intervals on the 40 Wilson coefficients considered in this work (in the linear
EFT case) between the outcome of the joint SMEFT-PDF determination and that of the fixed-PDF EFT analysis. The
latter is based on SM and EFT calculations performed with {\tt NNPDF4.0-notop} as input. In both cases,
results are based on the global dataset being considered and SMEFT cross-sections are evaluated up to linear,
corrections. The dashed horizontal line indicates the SM prediction, $c_k = 0$. Note that some coefficients are
multiplied by the indicated prefactor to facilitate the visualisation of the results.}
\label{fig:fixed_vs_simu}
\end{figure}

The first observation is that, similarly to the top-sector results observed in Ref.~\cite{Kassabov:2023hbm}, 
in the global fit presented here the interplay between PDFs and SMEFT Wilson coefficients is weak. 
Indeed in Fig.~\ref{fig:fixed_vs_simu} we observe that, including all data listed in Sect.~\ref{subsec:data}, 
the bounds on the SMEFT Wilson coefficients are essentially identical in a SMEFT-only fit and  
in a simultaneous fit of SMEFT WCs and PDFs. The only mild sign of interplay is shown by the 
operator $c_{\phi t}$, which undergoes a fair broadening in the simultaneous fit compared to the fixed-PDF 
SMEFT fit.

The PDF fit is more interesting. It was shown in Ref.~\cite{Kassabov:2023hbm} that when comparing: 
(i) a SM PDF fit excluding top data, (ii) a simultaneous PDF-SMEFT using top data, 
(iii) a SM PDF fit including top data, there is a hierarchy of shifts in the gluon-gluon luminosity 
at high invariant masses. In particular, the gluon-gluon luminosity of the simultaneous fit (ii) is reduced at high 
invariant masses compared to fit (i), whilst the gluon-gluon luminosity of the SM fit including all top data 
(iii) is even further reduced at high invariant masses relative to the result of the simultaneous fit (ii). 
This can be explained due to the additional SMEFT degrees of freedom in (ii) allowing for a better description 
of the top data with the PDF remaining compatible with the no-top PDF.

\begin{figure}[tb!]
\centering
\includegraphics[width=0.7\textwidth]{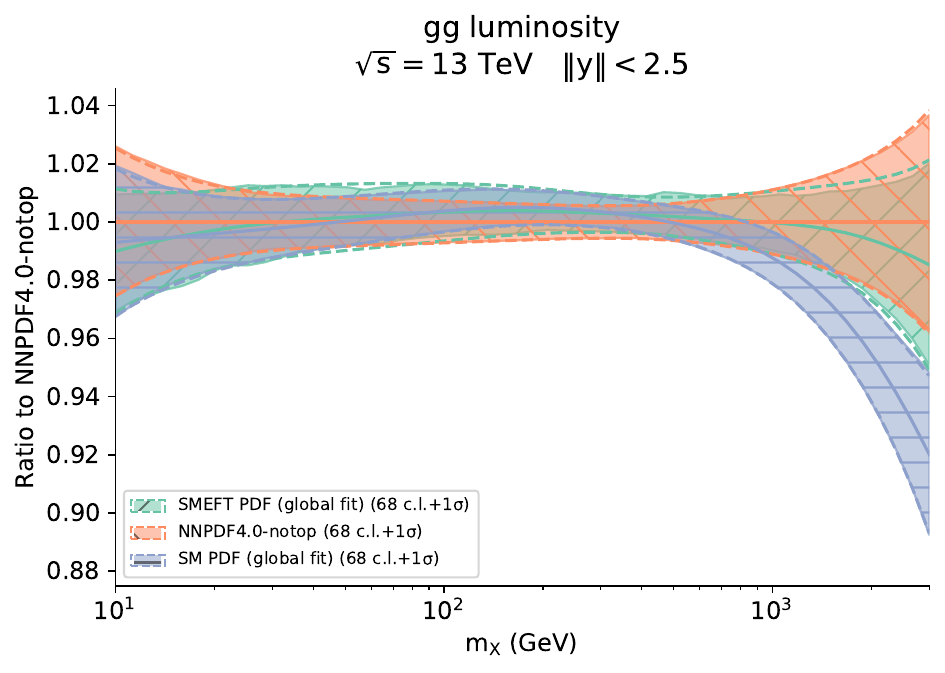}
\caption{ \label{fig:gglumi} The gluon-gluon partonic luminosities at $\sqrt{s} = 13$ TeV
as a function of the final-state invariant mass $m_X$. We compare the {\tt NNPDF4.0-notop} 
baseline fit with its SM-PDF counterpart including all new data presented in Sect.~\ref{subsec:data}, 
as well as with the simultaneous SMEFT-PDF determination. Results are presented as the ratio to the 
central value of the {\tt NNPDF4.0-notop} baseline.}
\end{figure}
As it can be observed in Fig.~\ref{fig:gglumi}, 
in the new global fit presented in this paper, there is a shift in the gluon-gluon luminosity in the simultaneous 
PDF and SMEFT fit relative to the baseline no-top PDF set, however the shift is much smaller than the shift that one 
has by including all data in the SM, meaning that the effect of the inclusion of the top, Higgs, EWPO and diboson
data has a much smaller effect on the PDFs due to the interplay between the pull of the SMEFT coefficients and the 
pull of the new data on the PDFs.  
The shift is even smaller compared to the shift of the simultaneous top fit relative to the no-top 
PDF presented in Ref.~\cite{Kassabov:2023hbm}. 
This is due to the inclusion of new high-mass Drell-Yan data in the new fit, which were not 
included in \nnpdf{}. These data favour a softer singlet PDF combination in the high-mass region, 
and as a result they enhance the gluon in the same region, due to sum rules. 

Finally, it is also interesting to compare the fit quality of the SMEFT fit keeping PDF fixed and the
simultaneous SMEFT-PDF fit. We display the $\chi^2$ per datapoint in Tables~\ref{tab:chi2-top-pair},~\ref{tab:chi2-single-top} and ~\ref{tab:chi2-other}
for each dataset appearing in both the simultaneous SMEFT-PDF fit and the fixed PDF SMEFT fit.
The fit quality for inclusive and associated top pair production is shown in Table~\ref{tab:chi2-top-pair},
for inclusive and associated single top production in Table~\ref{tab:chi2-single-top} and
for Higgs, diboson and electroweak precision observables in Table~\ref{tab:chi2-other}.
We also display the total fit quality of the 366 SMEFT-affected datapoints in Table~\ref{tab:chi2-other}.

Overall, we observe that the fit quality of the total dataset remains stable between the SMEFT-PDF
fit and the SMEFT fit, with a total $\chi^2$ of 0.91 for both fits, as shown in Table~\ref{tab:chi2-other}.
This is also observed in the Higgs and electroweak precision observables, while in the diboson sector
a small improvement from $\chi^2 = 1.24$ to $\chi^2 = 1.23$
is observed.
A similar improvement in the $\chi^2$ is observed in the inclusive top quark pair production 
datasets in Tables~\ref{tab:chi2-top-pair} from $\chi^2=1.03$ to $\chi^2=1.01$, and in inclusive single top datasets 
in Tables~\ref{tab:chi2-single-top} from $\chi^2=0.34$ to $\chi^2=0.33$, indicating that
in these sectors, the SMEFT-PDF fit provides a better fit to the data.  This improvement in 
the $\chi^2$ in these sectors is not as significant as the improvement observed in Ref.~\cite{Kassabov:2023hbm},
however.  As discussed below Fig.~\ref{fig:gglumi}, this is due to the inclusion of new high-mass Drell-Yan data in the new fit.

%%%%%%%%%%%%%%%%%%%%%%%%%%%%%%%%%%%%%%%%%%%%%%%%%%%%%%%%%%%%%%%%%%%%%%
\begin{table}[htbp]
    \begin{center}
    \renewcommand{\arraystretch}{1.60}
  %\scriptsize
    \tiny
  \begin{tabular}{ l | c| C{1.4cm} |  C{1.4cm} }
   \toprule
   Dataset & $n_{\rm dat}$   &  \multicolumn{2}{c}{$\chi^2/n_{\rm dat}$}  \\[1.5ex]
    &   & SMEFT-PDF fit & SMEFT fit   \\
   \midrule
      ATLAS $\sigma(t\bar{t})$, dilepton, 7 TeV    & 1    & 1.87 &  1.84 \\
      ATLAS $\sigma(t\bar{t})$, dilepton, 8 TeV & 1 & 0.33 & 0.35 \\
      ATLAS $1/\sigma d\sigma/dm_{t\bar{t}}$, dilepton, 8 TeV & 5 & 0.30 &  0.30 \\
      ATLAS $\sigma(t\bar{t})$, $\ell+$jets, 8 TeV & 1 & 3$\cdot 10^{-5}$ &  3$\cdot 10^{-4}$ \\
      ATLAS $1/\sigma d\sigma/d|y_t|$, $\ell+$jets, 8 TeV & 4&  1.27 & 1.22 \\
      ATLAS $1/\sigma d\sigma/d|y_{t\bar{t}}|$, $\ell+$jets, 8 TeV & 4 & 2.8 & 2.95  \\
      ATLAS $\sigma(t\bar{t})$, dilepton, 13 TeV           & 1   & 2$\cdot 10^{-3}$ & 2$\cdot 10^{-4}$  \\
      ATLAS $\sigma(t\bar{t})$, hadronic, 13 TeV         & 1    & 0.09 &  0.09   \\
      ATLAS $1/\sigma d^2\sigma/d|y_{t\bar{t}}|dm_{t\bar{t}}$, hadronic, 13 TeV & 10 & 1.84 & 1.88 \\
      ATLAS $\sigma(t\bar{t})$, $\ell+$jets, 13 TeV & 1 & 0.01 & 0.01 \\
      ATLAS $1/\sigma d\sigma/dm_{t\bar{t}}$, $\ell+$jets, 13 TeV & 8 & 2.07 & 2.1  \\
      \midrule
      CMS $\sigma(t\bar{t})$, combined, 5 TeV & 1 & 0.23 &  0.22 \\
      CMS $\sigma(t\bar{t})$, combined, 7 TeV & 1 & 0.01 & 0.01  \\
      CMS $\sigma(t\bar{t})$, combined, 8 TeV & 1 & 0.12 & 0.13  \\
      CMS $1/\sigma d^2\sigma/d|y_{t\bar{t}}|dm_{t\bar{t}}$, dilepton, 8 TeV & 16 & 0.51 & 0.53  \\
      CMS $1/\sigma d\sigma/d|y_{t\bar{t}}|$, $\ell+$jets, 8 TeV & 9  & 0.93 & 0.94  \\
      CMS $\sigma(t\bar{t})$, dilepton, 13 TeV & 1 & 0.60 & 0.62 \\
      CMS $1/\sigma d\sigma/dm_{t\bar{t}}$, dilepton, 13 TeV & 5 & 2.26 & 2.24  \\
      CMS $\sigma(t\bar{t})$, $\ell+$jets, 13 TeV & 1 & 1.90 & 1.98 \\
      CMS $1/\sigma d\sigma/dm_{t\bar{t}}$, $\ell+$jets, 13 TeV & 14& 0.93 & 0.94  \\
      \midrule
      ATLAS charge asymmetry, 8 TeV & 1 & 0.63 & 0.63\\
      ATLAS charge asymmetry, 13 TeV & 5& 0.87 & 0.91 \\
      CMS charge asymmetry, 8 TeV & 3 & 0.06 & 0.06\\
      CMS charge asymmetry, 13 TeV & 3 & 0.39 & 0.36  \\
      ATLAS \& CMS combined charge asy., 8 TeV & 6 & 0.60 & 0.61\\
      \midrule
      ATLAS $W$-hel., 13 TeV & 2 & 7$\cdot 10^{-5}$ &  1$\cdot 10^{-3}$ \\
      ATLAS \& CMS combined $W$-hel., 8 TeV & 2&  0.33 & 0.34  \\
         \midrule
	  {\bf Total inclusive $t\bar{t}$}  & {\bf 108}  & {\bf 1.01} & {\bf 1.03}  \\
       \midrule
       \midrule
      ATLAS $\sigma(t\bar{t}Z)$, 8 TeV & 1 & 1.40 & 1.33 \\
      ATLAS $\sigma(t\bar{t}W)$, 8 TeV & 1 & 0.62 & 0.71   \\
      ATLAS $\sigma(t\bar{t}Z)$, 13 TeV & 1 & 2$\cdot 10^{-6}$ & 5$\cdot 10^{-3}$  \\
      ATLAS $1/\sigma d\sigma(t\bar{t}Z)/dp_T^Z$, 13 TeV & 5 & 1.85 & 1.84 \\
      ATLAS $\sigma(t\bar{t}W)$, 13 TeV & 1& 2$\cdot 10^{-2}$ & 4$\cdot 10^{-3}$\\
      \midrule
      ATLAS $\sigma(t\bar{t}\gamma)$, 8 TeV & 1 & 0.33 & 0.29  \\
      \midrule
      CMS $\sigma(t\bar{t}Z)$, 8 TeV & 1 & 2$\cdot 10^{-3}$ & 5$\cdot 10^{-3}$\\
      CMS $\sigma(t\bar{t}W)$, 8 TeV & 1 & 0.69 & 0.78  \\
      CMS $\sigma(t\bar{t}Z)$, 13 TeV & 1& 0.10 & 0.15 \\
      CMS $1/\sigma d\sigma(t\bar{t}Z)/dp_T^Z$, 13 TeV & 3 & 0.86 & 0.88 \\
      CMS $\sigma(t\bar{t}W)$, 13 TeV & 1 & 0.48 & 0.35 \\
      \midrule
      CMS $\sigma(t\bar{t}\gamma)$, 8 TeV & 1 & 0.01 & 7$\cdot 10^{-3}$ \\
      \midrule
     {\bf Total associated $t\bar{t}$}  & {\bf 18}  &  {\bf 0.86} &  {\bf 0.86 }  \\
   \bottomrule
  \end{tabular}
  \end{center}
    \caption{\small \label{tab:chi2-top-pair}
    The values of the $\chi^2$ per data point for the datasets that enter both the SMEFT-only fit and the 
    simultaneous SMEFT-PDF fit. For each dataset, we indicate the number of data points $n_{\rm dat}$ and the 
    $\chi^2$ for the SMEFT-only fit with PDF fixed to the input PDF {\tt NNPDF4.0-notop}  and for the 
    simultaneous SMEFT-PDF determination. 
  }
  \end{table}
  %%%%%%%%%%%%%%%%%%%%%%%%%%%%%%%%%%%%%%%%%%%%%%%%%%%%%%%%%%%%%%%%%%%%%%%%%%
  %%%%%%%%%%%%%%%%%%%%%%%%%%%%%%%%%%%%%%%%%%%%%%%%%%%%%%%%%%%%%%%%%%%%%%%%%%%%%%%%

%%%%%%%%%%%%%%%%%%%%%%%%%%%%%%%%%%%%%%%%%%%%%%%%%%%%%%%%%%%%%%%%%%%%%%
\begin{table}[htbp]
    \begin{center}
    \renewcommand{\arraystretch}{1.60}
  %\scriptsize
    \tiny
  \begin{tabular}{ l | c| C{1.4cm} |  C{1.4cm} }
   \toprule
   Dataset & $n_{\rm dat}$   &  \multicolumn{2}{c}{$\chi^2/n_{\rm dat}$}  \\[1.5ex]
    &   & SMEFT-PDF fit & SMEFT fit   \\
   \midrule 
	  ATLAS $t$-channel $\sigma(t)$, 7 TeV & 1 & 0.24 & 0.26 \\
	  ATLAS $t$-channel $\sigma(\bar{t})$, 7 TeV & 1 & 0.01 & 0.02 \\
	  ATLAS $t$-channel $1/\sigma d\sigma(tq)/dy_t$, 7 TeV & 3 & 0.9 & 0.89 \\
	  ATLAS $t$-channel $1/\sigma d\sigma(\bar{t}q)/dy_{\bar{t}}$, 7 TeV & 3 & 0.06 & 0.06 \\
	  ATLAS $t$-channel $\sigma(t)$, 8 TeV & 1 & 0.01 & 0.02 \\
	  ATLAS $t$-channel $\sigma(\bar{t})$, 8 TeV & 1 & 0.30 & 0.24 \\
	  ATLAS $t$-channel $1/\sigma d\sigma(tq)/dy_t$, 8 TeV & 3 & 0.28 & 0.28 \\
	  ATLAS $t$-channel $1/\sigma d\sigma(\bar{t}q)/dy_{\bar{t}}$, 8 TeV & 3 & 0.19 & 0.20 \\
	  ATLAS $s$-channel $\sigma(t + \bar{t})$, 8 TeV & 1 & 0.02 & 0.04 \\
	  ATLAS $t$-channel $\sigma(t)$, 13 TeV & 1 & 0.31 & 0.32 \\
	  ATLAS $t$-channel $\sigma(\bar{t})$, 13 TeV & 1 & 0.05 & 0.03 \\
	  ATLAS $s$-channel $\sigma(t + \bar{t})$, 13 TeV & 1 & 0.05 & 0.03 \\
          \midrule
	  CMS $t$-channel $\sigma(t) + \sigma(\bar{t})$, 7 TeV & 1 & 0.02 & 0.02 \\
	  CMS $t$-channel $\sigma(t)$, 8 TeV & 1 & 0.34 & 0.31 \\
	  CMS $t$-channel $\sigma(\bar{t})$, 8 TeV & 1 & 0.98 & 1.07 \\
	  CMS $s$-channel $\sigma(t + \bar{t})$, 8 TeV & 1 & 1.42 & 1.45 \\
	  CMS $t$-channel $\sigma(t)$, 13 TeV & 1 & 0.34 & 0.35 \\
	  CMS $t$-channel $\sigma(\bar{t})$, 13 TeV & 1 & 0.02 & 0.03 \\
	  CMS $t$-channel $1/\sigma d\sigma/d|y^{(t)}|$, 13 TeV & 4 & 0.40 & 0.40 \\
         \midrule
	  {\bf Total inclusive single top}  & {\bf 30}  & {\bf 0.33} & {\bf 0.34}  \\
         \midrule
         \midrule
	  ATLAS $\sigma(tW)$, dilepton, 8 TeV & 1 & 0.07 & 0.05 \\
	  ATLAS $\sigma(tW)$, single-lepton, 8 TeV & 1 & 0.35 & 0.32 \\
	  ATLAS $\sigma(tW)$, dilepton, 13 TeV & 1 & 0.74 & 0.71 \\
	  ATLAS $\sigma_{\text{fid}}(tZj)$, dilepton, 13 TeV & 1 & 0.30 & 0.24 \\
	 \midrule
	  CMS $\sigma(tW)$, dilepton, 8 TeV & 1 & 0.08 & 0.07 \\
	  CMS $\sigma(tW)$, dilepton, 13 TeV & 1 & 2.21 & 2.41 \\
	  CMS $\sigma_{\text{fid}}(tZj)$, dilepton, 13 TeV & 1 & 0.13 & 0.17 \\
	  CMS $d\sigma_{\text{fid}}(tZj)/dp_T^{t}$, dilepton, 13 TeV & 3 & 0.13 & 0.14 \\
	  CMS $\sigma(tW)$, single-lepton, 13 TeV & 1 & 1.51 & 1.42 \\
	 \midrule
         {\bf Total associated single top}  & {\bf 11}  &  {\bf 0.53} &  {\bf 0.53}  \\
	 \bottomrule
  \end{tabular}
  \end{center}
    \caption{\small \label{tab:chi2-single-top}
	Same as Tab.~\ref{tab:chi2-top-pair}, now for inclusive and associated single top datasets.}
  \end{table}
  %%%%%%%%%%%%%%%%%%%%%%%%%%%%%%%%%%%%%%%%%%%%%%%%%%%%%%%%%%%%%%%%%%%%%%%%%%
  %%%%%%%%%%%%%%%%%%%%%%%%%%%%%%%%%%%%%%%%%%%%%%%%%%%%%%%%%%%%%%%%%%%%%%%%%%%%%%%%

%%%%%%%%%%%%%%%%%%%%%%%%%%%%%%%%%%%%%%%%%%%%%%%%%%%%%%%%%%%%%%%%%%%%%%
\begin{table}[htbp]
    \begin{center}
    \renewcommand{\arraystretch}{1.60}
  %\scriptsize
    \tiny
  \begin{tabular}{ l | c| C{1.4cm} |  C{1.4cm} }
   \toprule
   Dataset & $n_{\rm dat}$   &  \multicolumn{2}{c}{$\chi^2/n_{\rm dat}$}  \\[1.5ex]
    &   & SMEFT-PDF fit & SMEFT fit   \\
   \midrule 
	 ATLAS STXS combination, $\mu_{H}$, 13 TeV & 25 & 0.49 & 0.50 \\
	 ATLAS signal strength $H \rightarrow Z \gamma$, 13 TeV & 1 & 4$\cdot 10^{-5}$ & 3$\cdot 10^{-3}$ \\
	 ATLAS signal strength $H \rightarrow \mu^{+} \mu^{-}$, 13 TeV & 1 & 7$\cdot 10^{-3}$ & 2$\cdot 10^{-3}$ \\
	 CMS signal strength combination, $\mu_{H}$, 13 TeV & 24 & 0.66 & 0.67 \\
	 ATLAS \& CMS signal strength combination, $\mu_{H}$,  7 and 8 TeV & 22 & 0.98 & 0.96 \\
	 \midrule
	 {\bf Total Higgs}  & {\bf 73}  &  {\bf 0.68} &  {\bf 0.68}  \\
	 \midrule
	 \midrule
	 ATLAS $d \sigma _{W^+W^-}/d m_{e \mu}$, 13 TeV & 13 & 1.69 & 1.70 \\
	 ATLAS $d \sigma _{WZ} / d m_{T}$, 13 TeV & 6 & 0.77 & 0.77 \\
	 ATLAS $d \sigma(Zjj)/d \Delta \phi_{jj}$, 13 TeV & 12 & 0.79 & 0.79 \\
	 CMS $d \sigma _{WZ} / d p_{T}$, 13 TeV & 11 & 1.42 & 1.43 \\
	 LEP $d \sigma _{WW} / d cos(\theta _W)$, 0.182 TeV & 10 & 1.39 & 1.39 \\
	 LEP $d \sigma _{WW} / d cos(\theta _W)$, 0.189 TeV & 10 & 0.91 & 0.92 \\
	 LEP $d \sigma _{WW} / d cos(\theta _W)$, 0.198 TeV & 10 & 1.58 & 1.57 \\
         LEP $d \sigma _{WW} / d cos(\theta _W)$, 0.206 TeV & 10 & 1.07 & 1.08 \\
	 \midrule
	  {\bf Total Diboson}  & {\bf 82}  &  {\bf 1.23} &  {\bf 1.24 }  \\
	 \midrule 
	 \midrule
	 LEP $Z$ observables, 0.25 TeV & 19 & 0.52  & 0.52 \\
	 LEP $\mathcal{B}(W \rightarrow l^{-} \bar{v}_l)$ 0.196 TeV & 3 & 2.59 & 2.59 \\
	 LEP $\sigma(e^+ e^- \rightarrow e^+ e^-)$ 0.189 TeV & 21 & 1.02 & 1.03 \\
	 LEP $\hat{\alpha}^{(5)}_{\rm}(M_Z)$ 0.209 TeV & 1 & 0.47 & 0.25 \\
	 \midrule
	 {\bf Total EWPO}  & {\bf 44}  &  {\bf 0.90} &  {\bf 0.90 }  \\
	  \midrule
	  \midrule
	 {\bf Total} & {\bf 366} & {\bf 0.91} & {\bf 0.91} \\
	 \bottomrule
  \end{tabular}
  \end{center}
	\caption{\small \label{tab:chi2-other}
	Same as Tab.~\ref{tab:chi2-top-pair}, now for the Higgs, diboson and electroweak precision observables.}
  \end{table}
  %%%%%%%%%%%%%%%%%%%%%%%%%%%%%%%%%%%%%%%%%%%%%%%%%%%%%%%%%%%%%%%%%%%%%%%%%%
  %%%%%%%%%%%%%%%%%%%%%%%%%%%%%%%%%%%%%%%%%%%%%%%%%%%%%%%%%%%%%%%%%%%%%%%%%%%%%%%%

%To include
%\begin{itemize}  
%\item distribution of individual SMEFT coefficients?
%\item correlation between the Wilson coefficients and the PDFs.
%\simunet{} calculates the PDF-EFT correlation coefficient for a Wilson coefficient $c$ and a
%PDF $f(x, Q)$ at a given $x$ and $Q^2$ as
%\begin{equation}
%\label{eq:correlationL2CT}
%\rho\lp c, f(x,Q^2)\rp=\frac{\la c^{(k)}f^{(k)}(x,Q^2)\ra_k - \la c^{(k)}\ra_k \la f^{(k)}(x,Q^2)\ra_k
%}{\sqrt{\la (c^{(k)})^2\ra_k - \la c^{(k)}\ra_k^2}\sqrt{\la\left( f^{(k)}(x,Q^2)\right)^2\ra_k - \la f^{(k)}(x,Q^2)\ra_k^2}}  \, ,
%\end{equation}
%where $c^{(k)}$ is the best-fit value of the Wilson coefficient for
%the $k$-th replica and $f^{(k)}$ is the $k$-th PDF replica computed at
%a given $x$ and $Q$, and $\la \cdot \ra_k$ represents the average
%over all replicas. 
%

\section{Conclusions}
\label{sec:conclusions}

In this work we have presented the public release, as an open-source
code, of the software framework underlying the \simunet{} 
global simultaneous determination of PDFs and SMEFT coefficients. 
Based on the first tagged version of the public {\tt NNPDF} code~\cite{nnpdfcode,NNPDF:2021uiq}, 
\simunet{} adds a number of new
features that allow the exploration of the interplay between a fit of PDFs and 
BSM degrees of freedom.

We have documented the code and shown the phenomenological studies that can be performed with it, 
that include (i) a PDF-only fit {\it \`{a} la} \nnpdf{} with extra datasets included;
(ii) a global EFT analysis of the full top quark sector together with the Higgs production and decay
  rates data from the LHC, and the precision electroweak and diboson measurements
  from LEP and the LHC; (iii) a simultaneous PDF-SMEFT fit in which the user can decide which 
  observable to treat as PDF-dependent or independent by either fitting PDFs alongside the Wilson
  coefficients or freezing them to a baseline PDF set.

We have demonstrated that the user can use the code to inject any New Physics model 
in the data and check whether the model can be fitted away by the PDF parametrisation. 
We have also shown that the methodology successfully passes the closure test with 
respect to an underlying PDF law and UV "true" model that is injected in the data. 

The new result presented here is the global analysis of the SMEFT and explicitly shows that 
\simunet{} can be used to produce fits combining the Higgs, top, diboson and 
electroweak sectors such as those in Refs.~\cite{Ellis:2020unq,Ethier:2021bye,Giani:2023gfq}, 
and that more data can be added in a rather straightforward way to combine the sectors presented here alongside Drell-Yan 
and flavour observables, as it is done in Ref.~\cite{Bartocci:2023nvp}. 
We have presented the results and compared them to existing fits. We have shown that the interplay between PDFs 
and SMEFT coefficients, once they are fitted simultaneously, has little effect on the Wilson coefficients, while 
it has the effect of shifting the gluon-gluon luminosity and increasing its uncertainty in the large-$M_X$ region. 

The public release of \simunet{} represents a first crucial step towards the interpretation 
of indirect searches under a unified framework. We can assess the impact of new
datasets not only on the PDFs, but now on the couplings of an EFT expansion, or on any
other physical parameter.
Indeed, the methodology presented here can be extended to simultaneous fitting
precision SM parameters, such as the strong coupling or electroweak parameters. Indeed,
our framework extends naturally, not only to BSM studies, but to any parameter which
may modify the SM prediction through the use of K-factors or other kind of interpolation, see
Appendix A of Ref.~\cite{Iranipour:2022iak} for the application of \simunet{} to the
simultaneous fit of PDFs and $\alpha_s(M_z)$~\cite{Forte:2020pyp}.

The \simunet{} code is publicly available from its {\sc\small GitHub} repository:
\begin{center}
  {\tt \url{https://github.com/HEP-PBSP/SIMUnet}},
\end{center}
and is accompanied by documentation and tutorials provided at:
\begin{center}
 \web{}.
\end{center}

\section*{Acknowledgements}
We thank Shayan Iranipour for his precious contributions in building
\simunet{}. We thank the \nnpdf{} collaboration for having made their
code available, so that new phenomenological directions such as the
one that we presented here, can be explore. We thank the \smefit{} collaboration for having shared the
SMEFT theoretical predictions and data implementation of some of the 
new observables that we include. 
Mark N.~Costantini, Elie Hammou, Zahari Kassabov, Luca Mantani, Manuel
Morales-Alvarado, James M.~ Moore and Maria Ubiali are supported
by the European Research Council under the European Union’s
Horizon 2020 research and innovation Programme (grant agreement
n.950246), and partially by the STFC
consolidated grant ST/T000694/1 and ST/X000664/1.
The work of Maeve Madigan is supported by the Deutsche Forschungsgemeinschaft (DFG, German
Research Foundation) under grant 396021762 – TRR 257 Particle Physics Phenomenology
after the Higgs Discovery.
%------------------------------------------------------

%\bibliography{simucode}

\end{document}